\documentclass[aps,prd,twocolumn,superscriptaddress]{revtex4}
\usepackage{epsfig,epsf}
\usepackage{amsmath}
\usepackage{amsthm}
\usepackage{amsfonts}
\usepackage{amssymb}
\usepackage{dsfont}
\usepackage{multirow}
\usepackage{appendix}
\usepackage{slashed}
\usepackage[active]{srcltx}
\usepackage{psfrag}

\begin{document}

\title{Fully charmed resonance $X(6900)$ and its beauty counterpart }
\date{\today}
\author{S.~S.~Agaev}
\affiliation{Institute for Physical Problems, Baku State University, Az--1148 Baku,
Azerbaijan}
\author{K.~Azizi}
\affiliation{Department of Physics, University of Tehran, North Karegar Avenue, Tehran
14395-547, Iran}
\affiliation{Department of Physics, Do\v{g}u\c{s} University, Dudullu-\"{U}mraniye, 34775
Istanbul, T\"{u}rkiye}
\author{B.~Barsbay}
\affiliation{Division of Optometry, School of Medical Services and Techniques, Do\v{g}u%
\c{s} University, 34775 Istanbul, T\"{u}rkiye}
\author{H.~Sundu}
\affiliation{Department of Physics Engineering, Istanbul Medeniyet University, 34700
Istanbul, T\"{u}rkiye}

\begin{abstract}
The fully heavy scalar tetraquarks $T_{\mathrm{4Q}}=QQ\overline{Q}\overline{Q%
}$, ($Q=c, b$) are explored in the context of QCD sum rule method. We model $%
T_{\mathrm{4Q}}$ as diquark-antidiquark systems composed of pseudoscalar
constituents, and calculate their masses $m^{(\prime)}$ and couplings $%
f^{(\prime)}$ within the two-point sum rule approach. Our results $m=(6928
\pm 50)~\mathrm{MeV}$ and $m^{\prime}=(18858 \pm 50)~\mathrm{MeV}$ for
masses of the tetraquarks $T_{\mathrm{4c}}$ and $T_{\mathrm{4b}}$ prove that
they can decay to hidden-flavor heavy mesons. The full width $\Gamma_{%
\mathrm{4c}}$ of the $T_{\mathrm{4c}}$ is evaluated by taking into account
the decay channels $T_{\mathrm{4c}} \to J/\psi J/\psi $, $J/\psi
\psi^{\prime}$, $\eta _{c}\eta _{c}$, $\eta _{c}\eta _{c}(2S)$, $\eta
_{c}\chi_{c1}(1P)$, and $\chi_{c0}\chi_{c0}$. The partial widths of these
processes depend on strong couplings $g_{i}$ at vertices $T_{\mathrm{4c}%
}J/\psi J/\psi $, $T_{\mathrm{4c}}J/\psi \psi^{\prime} $ etc., which are
computed using the QCD three-point sum rule method. The decay $T_{\mathrm{4b}%
} \to \eta_{b}\eta_{b}$ is used to find the width $\Gamma_{\mathrm{4b}}$ of
the $T_{\mathrm{4b}}$. The predictions for $m $ and $\Gamma_{\mathrm{4c}%
}=(128 \pm 22)~\mathrm{MeV}$ are compared with parameters of the fully
charmed resonances reported by the LHCb, ATLAS, and CMS Collaborations.
Based on this analysis, we interpret the tetraquark $T_{\mathrm{4c}}$ as a
candidate to the resonance $X(6900)$. The mass $m^{\prime} $ and width $%
\Gamma_{\mathrm{4b}}=(94 \pm 28)~\mathrm{MeV}$ of the exotic meson $T_{%
\mathrm{4b}}$ can be used in future experimental investigations of these
states.
\end{abstract}

\maketitle


\section{Introduction}

\label{sec:Int} 

It is known, that existence of exotic states composed of more than three
valence partons or bearing unusual quantum numbers is not forbidden by the
parton model and first principles of QCD. Phenomenological studies of such
multiquark hadrons started almost a half century ago from modeling the nonet
of light scalar mesons as $q^{2}\overline{q}^{2}$ four-quark states, and
analysis of the doubly strange hexaquark structure \cite%
{Jaffe:1976ig,Jaffe:1976yi}. Due to collected experimental information and
theoretical achievements multiquark hadrons are objects of intensive studies
in high energy physics.

Theoretical investigations of multiquark hadrons aimed to elaborate methods
to deal with such structures, calculate their parameters, and search for
processes to detect them. One of important problems of these studies was a
stability of multiquark hadrons against strong and/or electromagnetic
decays, because stable particles with long mean lifetimes can be easily
observed in various processes. Structures composed of heavy $QQ$ ($Q=c$ or $%
b $ ) diquarks and light antidiquarks are real candidates to such exotic
mesons. Thus, it was already demonstrated that compounds $QQ\overline{q}%
\overline{q}$ may be strong-interaction stable particles provided the ratio $%
m_{Q}/m_{q}$ is large \cite{Ader:1981db,Zouzou:1986qh,Lipkin:1986dw}. For
example, a conclusion about stable nature of the isoscalar axial-vector
tetraquark $bb\overline{u}\overline{d}$ was made in Ref.\ \cite%
{Carlson:1987hh} confirmed later by other researches \cite%
{Navarra:2007yw,Eichten:2017ffp,Karliner:2017qjm}. Evidently stable
tetraquarks transform to conventional mesons through weak processes, and
live considerably longer than other multiquark systems \cite%
{Xing:2018bqt,Agaev:2018khe,Li:2018bkh,Sundu:2019feu,Agaev:2019kkz,Agaev:2019lwh, Agaev:2020dba,Agaev:2020mqq,Agaev:2020zag,Yu:2017pmn}%
.

The heavy exotic mesons $QQ\overline{Q}^{(\prime )}\overline{Q}^{(\prime )}$
establish another class of particles, which deserves detailed theoretical
and experimental investigations. Recently, the LHCb, ATLAS and CMS\
Collaborations reported new structures discovered in di-$J/\psi $ and $%
J/\psi \psi ^{\prime }$ mass distributions \cite%
{LHCb:2020bwg,Bouhova-Thacker:2022vnt,CMS:2023owd}. The LHCb observed a
threshold enhancement in nonresonant di-$J/\psi $ production from $6.2$ to $%
6.8~\mathrm{GeV}$ with center at $6.49~\mathrm{GeV}$. A narrow peak $X(6900)$
at $6.9~\mathrm{GeV}$, and a resonance around $7.2~\mathrm{GeV}$ were seen
as well. The ATLAS and CMS experiments detailed information on structures in
the region $6.2-6.8~\mathrm{GeV}$ and at $7.2~\mathrm{GeV}$. Thus, ATLAS
detected the resonances $X(6200)$, $X(6600)$, and $X(6900)$ in the di-$%
J/\psi $ and $X(7300)$ in the $J/\psi \psi ^{\prime }$ channels,
respectively. The resonances $X(6600)$, $X(6900)$ and $X(7300)$ were fixed
by the CMS Collaboration as well.

Analyses performed in the context of various methods and models led to
interesting, sometimes to contradictory interpretations of the observed
resonances \cite%
{Zhang:2020xtb,Albuquerque:2020hio,Wang:2022xja,Dong:2022sef,
Faustov:2022mvs,Lu:2023ccs,Dong:2020nwy,Liang:2021fzr,Gong:2020bmg,Niu:2022vqp,Yu:2022lak,Kuang:2023vac}%
. In fact, $X(6900)$ was interpreted as a tetraquark built of pseudoscalar
diquark and antidiquark components \cite{Albuquerque:2020hio}. The $X(6200)$
was assigned to be the ground-level tetraquark state with $J^{\mathrm{PC}%
}=0^{++}$ or $1^{+-}$, whereas $X(6600)$ was considered as its first radial
excitation \cite{Wang:2022xja}. In Ref.\ \cite{Dong:2022sef} the authors
proposed to consider the resonances, starting from $X(6200)$, as the $1S$, $%
1P/2S$, $1D/2P$, and $2D/3P/4S$ tetraquark states. Similar interpretations
in the context of the relativistic quark model were suggested as well \cite%
{Faustov:2022mvs}. The resonance $X(6900)$ was modeled as hadronic molecules
$\chi _{c0}\chi _{c0}$ or $J/\psi \psi (3770)$,\ $\chi _{c0}\chi _{c2}$ in
Refs.\ \cite{Albuquerque:2020hio,Lu:2023ccs}, respectively.

Alternatively, in the framework of the coupled-channel approach a
near-threshold state in the di-$J/\psi $ system with $J^{\mathrm{PC}}=0^{++}$
or $J^{\mathrm{PC}}=2^{++}$ was interpreted as $X(6200)$ \cite{Dong:2020nwy}%
. Coupled-channel effects may also generate a pole structure identified in
Ref.\ \cite{Liang:2021fzr} with the resonance $X(6900)$. The $X(6900)$ may
be dynamically generated by the Pomeron exchanges and coupled-channel
effects between the $J/\psi J/\psi $ and $J/\psi \psi ^{\prime }$ scattering
\cite{Gong:2020bmg}.

In our article \cite{Agaev:2023wua}, we carried out rather detailed analysis
of the fully charmed $X_{\mathrm{4c}}$ and beauty $X_{\mathrm{4b}}$ scalar
four-quark mesons by calculating their masses and current couplings. We
modeled $X_{\mathrm{4c}}$ and $X_{\mathrm{4b}}$ as diquark-antidiquarks
built of axial-vector constituents $Q^{T}C\gamma _{\mu }Q$ and $\overline{Q}%
\gamma ^{\mu }C\overline{Q}^{T}$ (briefly, a tetraquark with a structure $%
C\gamma _{\mu }\otimes \gamma ^{\mu }C$), where $C$ is the charge
conjugation matrix. Our prediction for the mass $m=(6570\pm 55)~\mathrm{MeV}$
of the tetraquark $X_{\mathrm{4c}}$ proved that it can decay to $J/\psi
J/\psi $, $\ \eta _{c}\eta _{c}$, and $\eta _{c}\chi _{c1}(1P)$ mesons. The
full width of $X_{\mathrm{4c}}$ was computed using these decay modes and
found equal to $\Gamma _{\mathrm{4c}}=(110\pm 21)~\mathrm{MeV}$. Comparison
with available data allowed us to consider $X_{\mathrm{4c}}$ as a candidate
to the resonance $X(6600)$. We also argued that $X(7300)$ may be considered
as $2S$ excitation of the resonance $X(6600)$. In Ref. \cite{Agaev:2023wua},
we calculated the mass of the fully beauty scalar state $X_{\mathrm{4b}}$ as
well. It turned out that, its mass $m^{\prime }=(18540\pm 50)~\mathrm{MeV}$
is below the $\eta _{b}\eta _{b}$ threshold, and hence $X_{\mathrm{4b}}$ can
not decay to pairs of hidden-beauty mesons. Its decays run via a $\overline{b%
}b$ transformation to a gluon(s) and light quark-antiquark pairs followed by
creation of ordinary $B$ mesons \cite{Becchi:2020mjz,Becchi:2020uvq}. The
electroweak leptonic and nonleptonic decays of $X_{\mathrm{4b}}$ are also
among its possible transitions to conventional particles.

In the present work, we continue our explorations of the LHCb-ATLAS-CMS
resonances in the context of\ the diquark-antidiquark model. We compute
parameters of the scalar $J^{\mathrm{PC}}=0^{++}$ state $T_{\mathrm{4c}}$
built of pseudoscalar diquark components, i. e., a tetraquark with $C\otimes
C$ type internal organization. We calculate the mass and full width of $T_{%
\mathrm{4c}}$, and confront our results with the experimental data for the
fully charmed $X$ resonances. We investigate also its beauty counterpart $T_{%
\mathrm{4b}}$ to determine the mass and width of this state.

The current paper is organized in the following manner: In Section \ref%
{sec:Masses}, we calculate the mass and current coupling of the tetraquarks $%
T_{\mathrm{4c}}$ and $T_{\mathrm{4b}}$. In Sec.\ \ref{sec:Decays1}, we
consider decays of $T_{\mathrm{4c}}$ to mesons $J/\psi J/\psi $ and $J/\psi
\psi ^{\prime }$. The processes $T_{\mathrm{4c}}\rightarrow \eta _{c}\eta
_{c}$, $\eta _{c}\eta _{c}(2S)$ are analyzed in Sec.\ \ref{sec:Decays2}. The
decays $T_{\mathrm{4c}}\rightarrow \eta _{c}\chi _{c1}(1P)$ and $T_{\mathrm{%
4c}}\rightarrow \chi _{c0}\chi _{c0}$ are investigated in Sec.\ \ref%
{sec:Decays2A}. Here, we also determine the full width of $T_{\mathrm{4c}}$.
The width of the process $T_{\mathrm{4b}}\rightarrow \eta _{b}\eta _{b}$ is
computed in Sec.\ \ref{sec:Decays3}. Last section is reserved for discussion
of results and contains our concluding notes.


\section{Mass and current coupling of the tetraquarks $T_{\mathrm{4c}}$ and $%
T_{\mathrm{4b}}$}

\label{sec:Masses}

Here, we compute the masses $m^{(\prime )}$ and current couplings $%
f^{(\prime )}$ of the exotic fully heavy-flavor mesons $T_{\mathrm{4c}}$ and
$T_{\mathrm{4b}}$ in the context of the QCD two-point sum rule method \cite%
{Shifman:1978bx,Shifman:1978by}. To this end, we consider the correlation
function
\begin{equation}
\Pi (p)=i\int d^{4}xe^{ipx}\langle 0|\mathcal{T}\{J(x)J^{\dag
}(0)\}|0\rangle ,  \label{eq:CF1}
\end{equation}%
where, $\mathcal{T}$ is the time-ordered product of two currents, and $J(x) $
is the interpolating currents for these states. We treat $T_{\mathrm{4c}}$
and $T_{\mathrm{4b}}$ as tetraquarks built of pseudoscalar diquark $Q^{T}CQ$
and antidiquark $\overline{Q}C\overline{Q}^{T}$. Then relevant interpolating
current is defined by the expression
\begin{equation}
J(x)=Q_{a}^{T}(x)CQ_{b}(x)\overline{Q}_{a}(x)C\overline{Q}_{b}^{T}(x),
\label{eq:CR1}
\end{equation}%
where $a$, and $b$ are color indices, and $Q(x)$ is either $c$ or $b$ quark
field. The current $J(x)$ describes the diquark-antidiquark state with
quantum numbers $J^{\mathrm{PC}}=0^{++}$.

Below, we present formulas for the four-quark state $T_{\mathrm{4c}}$, which
can be readily extended to the case of $T_{\mathrm{4b}}$. The
phenomenological side of the sum rule $\Pi ^{\mathrm{Phys}}(p)$ can be found
from Eq.\ (\ref{eq:CF1}) by inserting a complete set of intermediate states
with quark content and spin-parities of the tetraquark $T_{\mathrm{4c}}$,
and carrying out integration over $x$. As a result, we get
\begin{equation}
\Pi ^{\mathrm{Phys}}(p)=\frac{\langle 0|J|T_{\mathrm{4c}}(p)\rangle \langle
T_{\mathrm{4c}}(p)|J^{\dagger }|0\rangle }{m^{2}-p^{2}}+\cdots ,
\label{eq:Phys1}
\end{equation}%
where $p$ is four-momentum of $T_{\mathrm{4c}}$. The dots in Eq.\ (\ref%
{eq:Phys1}) denote contributions of higher resonances and continuum states.

The correlation function $\Pi ^{\mathrm{Phys}}(p)$ can be rewritten in terms
of the tetraquark's parameters through
\begin{equation}
\langle 0|J|T_{\mathrm{4c}}(p)\rangle =fm,  \label{eq:ME1}
\end{equation}%
which gives
\begin{equation}
\Pi ^{\mathrm{Phys}}(p)=\frac{f^{2}m^{2}}{m^{2}-p^{2}}+\cdots .
\label{eq:Phen2}
\end{equation}%
The correlator $\Pi ^{\mathrm{Phys}}(p)$ has a Lorentz structure which is
proportional to $\mathrm{I}$. Then, an invariant amplitude $\Pi ^{\mathrm{%
Phys}}(p^{2})$ required to derive the sum rules is equal to right-hand side
of Eq.\ (\ref{eq:Phen2}).

The counterpart of $\Pi ^{\mathrm{Phys}}(p^{2})$, i.e., the invariant
amplitude $\Pi ^{\mathrm{OPE}}(p^{2})$ evaluated by employing the heavy
quark propagators establishes the QCD side of the sum rule. The function $%
\Pi ^{\mathrm{OPE}}(p^{2})$ is extracted from the correlator $\Pi ^{\mathrm{%
OPE}}(p)$ that is calculated in the operator product expansion ($\mathrm{OPE}
$) and contains only a component proportional to $\mathrm{I}$.

In terms of $c$-quark propagators $\Pi ^{\mathrm{OPE}}(p)$ is given by the
formula
\begin{eqnarray}
&&\Pi ^{\mathrm{OPE}}(p)=i\int d^{4}xe^{ipx}\left\{ \mathrm{Tr}\left[
\widetilde{S}_{c}^{b^{\prime }b}(-x)S_{c}^{a^{\prime }a}(-x)\right] \right.
\notag \\
&&\times \left[ \mathrm{Tr}\left[ \widetilde{S}_{c}^{aa^{\prime
}}(x)S_{c}^{bb^{\prime }}(x)\right] +\mathrm{Tr}\left[ \widetilde{S}%
_{c}^{ba^{\prime }}(x)\right. \right.  \notag \\
&&\left. \left. \times S_{c}^{ab^{\prime }}(x)\right] \right] +\mathrm{Tr}%
\left[ \widetilde{S}_{c}^{a^{\prime }b}(-x)S_{c}^{b^{\prime }a}(-x)\right]
\notag \\
&&\left. \times \left[ \mathrm{Tr}\left[ \widetilde{S}_{c}^{ba^{\prime
}}(x)S_{c}^{ab^{\prime }}(x)\right] +\mathrm{Tr}\left[ \widetilde{S}%
_{c}^{aa^{\prime }}(x)S_{c}^{bb^{\prime }}(x)\right] \right] \right\} .
\notag \\
&&  \label{eq:QCD1}
\end{eqnarray}%
Here,%
\begin{equation}
\widetilde{S}_{c}(x)=CS_{c}^{T}(x)C,  \label{eq:Prop}
\end{equation}%
and $S_{c}(x)$ is the $c$-quark propagator explicit expression of which can
be found in Appendix (see, also Refs.\ \cite{Agaev:2023wua,Agaev:2020zad}).

At the next stage of analysis, we equate the amplitudes $\Pi ^{\mathrm{Phys}%
}(p^{2})$ and $\Pi ^{\mathrm{OPE}}(p^{2}),$ apply the Borel transformation
to suppress effects of higher resonances and continuum states, and make use
of the assumption on quark-hadron duality to subtract these terms from the
physical side of the sum rule equality. After some trivial manipulations, we
derive the following sum rules for the mass $m$ and current coupling $f$ of
the tetraquark $T_{\mathrm{4c}}$
\begin{equation}
m^{2}=\frac{\Pi ^{\prime }(M^{2},s_{0})}{\Pi (M^{2},s_{0})}  \label{eq:Mass}
\end{equation}%
and
\begin{equation}
f^{2}=\frac{e^{m^{2}/M^{2}}}{m^{2}}\Pi (M^{2},s_{0}),  \label{eq:Coupl}
\end{equation}%
where $\Pi ^{\prime }(M^{2},s_{0})=d\Pi (M^{2},s_{0})/d(-1/M^{2})$. In Eqs.\
(\ref{eq:Mass}) and (\ref{eq:Coupl}) the function $\Pi (M^{2},s_{0})$ is the
amplitude $\Pi ^{\mathrm{OPE}}(p^{2})$ obtained after the Borel
transformation and continuum subtraction procedures. It depends on the Borel
$M^{2}$ and continuum subtraction $s_{0}$ parameters, which appear in the
sum rule equality after corresponding operations.

The function $\Pi (M^{2},s_{0})$ has the form%
\begin{equation}
\Pi (M^{2},s_{0})=\int_{16m_{c}^{2}}^{s_{0}}ds\rho ^{\mathrm{OPE}%
}(s)e^{-s/M^{2}},  \label{eq:InvAmp}
\end{equation}%
where $\rho ^{\mathrm{OPE}}(s)$ is a two-point spectral density determined
as an imaginary part of the invariant amplitude $\Pi ^{\mathrm{OPE}}(p^{2})$%
. In general, the operator production expansion for the correlation function
$\Pi ^{\mathrm{OPE}}(p)$, apart from the perturbative term, contains
contributions of gluon condensates $\sim \langle \alpha _{s}G^{2}/\pi
\rangle $, $\langle g_{s}^{3}G^{3}\rangle $ etc. Because effects of higher
dimensional terms are numerically small, we restrict ourselves by
considering only dimension-4 contribution, which is proportional to $\langle
\alpha _{s}G^{2}/\pi \rangle $. As a result, the function $\rho ^{\mathrm{OPE%
}}(s)$ consists of a perturbative term $\rho ^{\mathrm{pert.}}(s)$ and a
dimension-$4$ nonperturbative contribution $\sim \langle \alpha
_{s}G^{2}/\pi \rangle $%
\begin{equation}
\rho ^{\mathrm{OPE}}(s)=\rho ^{\mathrm{pert.}}(s)+\langle \frac{\alpha
_{s}G^{2}}{\pi }\rangle \rho ^{\mathrm{Dim4}}(s).  \label{eq:SD}
\end{equation}%
The explicit expression of $\rho ^{\mathrm{pert.}}(s)$ is written down in
Appendix. The function $\rho ^{\mathrm{Dim4}}(s)$ is rather lengthy and,
therefore, has not been provided there.

In the case under discussion, $\Pi ^{\mathrm{OPE}}(p)$ depends only the $c$%
-quark propagators, which does not contain the light quark and mixed
quark-gluon condensates. As a result, we need for numerical analysis of the
sum rules the masses of $c$ and $b$ quarks, as well as the gluon vacuum
condensate $\langle \alpha _{s}G^{2}/\pi \rangle $: Values of these
parameters are presented below%
\begin{eqnarray}
&&m_{c}=(1.27\pm 0.02)~\mathrm{GeV},  \notag \\
&&m_{b}=4.18_{-0.02}^{+0.03}~\mathrm{GeV},  \notag \\
&&\langle \frac{\alpha _{s}G^{2}}{\pi }\rangle =(0.012\pm 0.004)~\mathrm{GeV}%
^{4}.  \label{eq:Parameters}
\end{eqnarray}

One should also fix the working regions for parameters $M^{2}$ and $s_{0}$.
The $M^{2}$ and $s_{0}$ have to satisfy constraints imposed on $\Pi
(M^{2},s_{0})$ by a pole contribution ($\mathrm{PC}$) and convergence of the
operator product expansion. To evaluate $\mathrm{PC,}$ we employ the
expression%
\begin{equation}
\mathrm{PC}=\frac{\Pi (M^{2},s_{0})}{\Pi (M^{2},\infty )},  \label{eq:PC}
\end{equation}%
and require fulfillment of the constraint $\mathrm{PC}\geq 0.5$.

Because the sum rules for $m$ and $f$ depend only on a nonperturbative term $%
\sim \langle \alpha _{s}G^{2}/\pi \rangle $, the pole contribution becomes
an important criterium in the choice of $M^{2}$ and $s_{0}$. We demand also
a stability of extracted observables under variations of the Borel and
continuum subtraction parameters. In fact, $M^{2}$ and $s_{0}$ are the
auxiliary quantities in computations, therefore, the physical observables
should not depend on $M^{2}$ and $s_{0}$. In real analysis, however, $m$ and
$f$ bear such residual dependence. Hence, the maximal stability of $m$ and $%
f $ under variations of $M^{2}$ and $s_{0}$ is among important constraints
in the choice of relevant working intervals. The prevalence of a
perturbative contribution over nonperturbative one is a required condition
as well.

\begin{figure}[h]
\includegraphics[width=8.5cm]{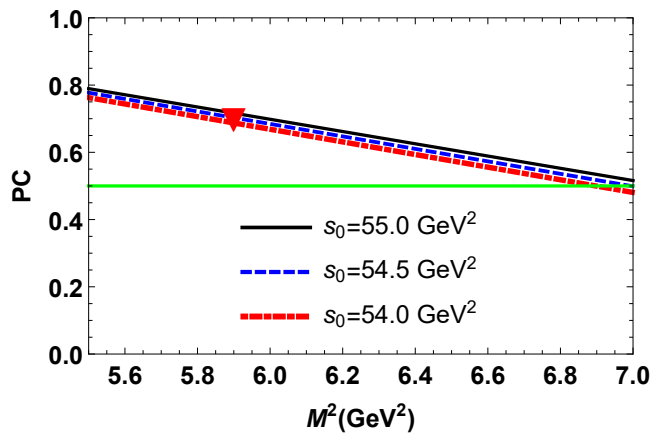}
\caption{The pole contribution $\mathrm{PC}$ as a function of the Borel
parameter $M^{2}$ at different $s_{0}$. The limit $\mathrm{PC}=0.5$ is
plotted by the horizontal line. The red triangle shows the point, where the
mass $m$ of the tetraquark $T_{\mathrm{4c}}$ has been extracted from the sum
rule. }
\label{fig:PC}
\end{figure}

\begin{figure}[h]
\includegraphics[width=8.5cm]{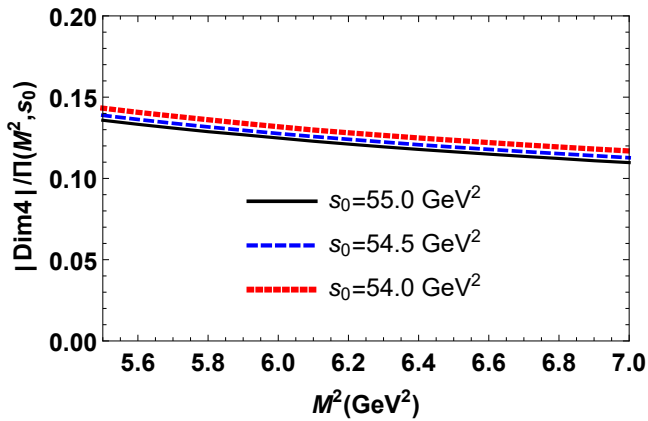}
\caption{The ratio of the nonperturbative $|\Pi^{ \mathrm{Dim4}}(M^2,s_0)|$
contribution and the whole $\Pi(M^2,s_0)$ correlation function as a function
of the Borel parameter at fixed $s_0$.}
\label{fig:Ratio}
\end{figure}

By employing $\mathrm{PC}$ one can fix the higher limit of the Borel
parameter $M^{2}$. The lower border for $M^{2}$ is found from a stability of
the sum rules' results under variation of $M^{2}$, and from superiority of
the perturbative term in extracted quantities. Two values of $M^{2}$
determined by this manner limit the region where $M^{2}$ can be changed.
Numerical computations demonstrate that the regions
\begin{equation}
M^{2}\in \lbrack 5.5,7]~\mathrm{GeV}^{2},\ s_{0}\in \lbrack 54,55]~\mathrm{%
GeV}^{2},  \label{eq:Wind1}
\end{equation}%
meet all necessary constraints imposed on the correlation function by the
sum rule analyses. Thus, at $M^{2}=5.5~\mathrm{GeV}^{2}$ and $M^{2}=7~%
\mathrm{GeV}^{2}$ the pole contribution equals to $0.78$ and $0.51$,
respectively. In order to show dynamics of the pole contribution, in Fig.\ %
\ref{fig:PC} we depict $\mathrm{PC}$ as a function of $M^{2}$ at different $%
s_{0}$: It exceeds $0.5$ for all values of the parameters $M^{2}$ and $s_{0}$
from Eq.\ (\ref{eq:Wind1}).

The nonperturbative term $\Pi ^{\mathrm{Dim4}}(M^{2},s_{0})$ is negative,
and at the minimum $M^{2}=5.5~\mathrm{GeV}^{2}$ forms $14\%$ of the
correlation function gradually decreasing with $M^{2}$ (see, Fig.\ \ref%
{fig:Ratio}). The next term $\sim \langle g_{s}^{3}G^{3}\rangle $ in $%
\mathrm{OPE}$ would be considerably smaller than the dimension-4
contribution and is neglected almost in all sum rule analysis of fully heavy
tetraquarks.

\begin{widetext}

\begin{figure}[h!]
\begin{center}
\includegraphics[totalheight=6cm,width=8cm]{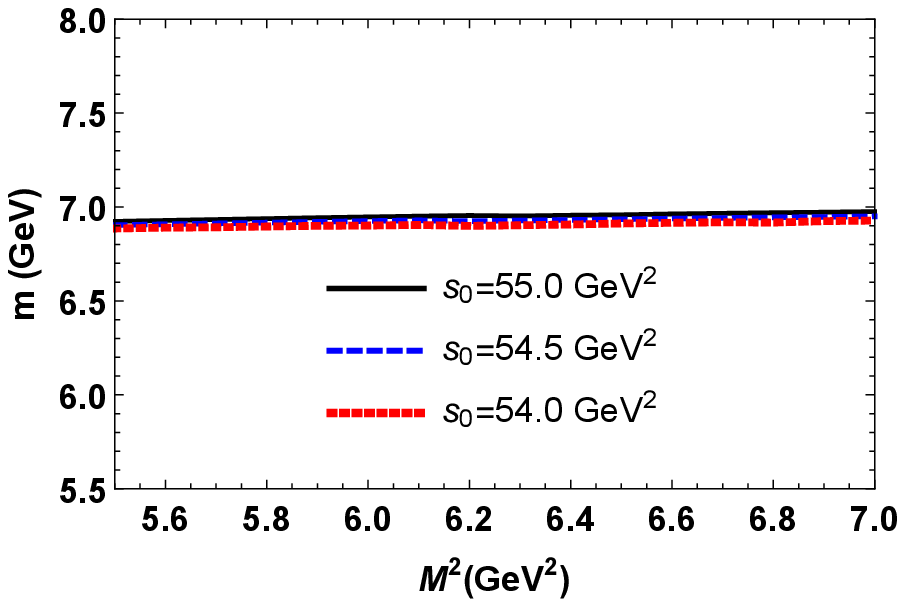}\,\, %
\includegraphics[totalheight=6cm,width=8cm]{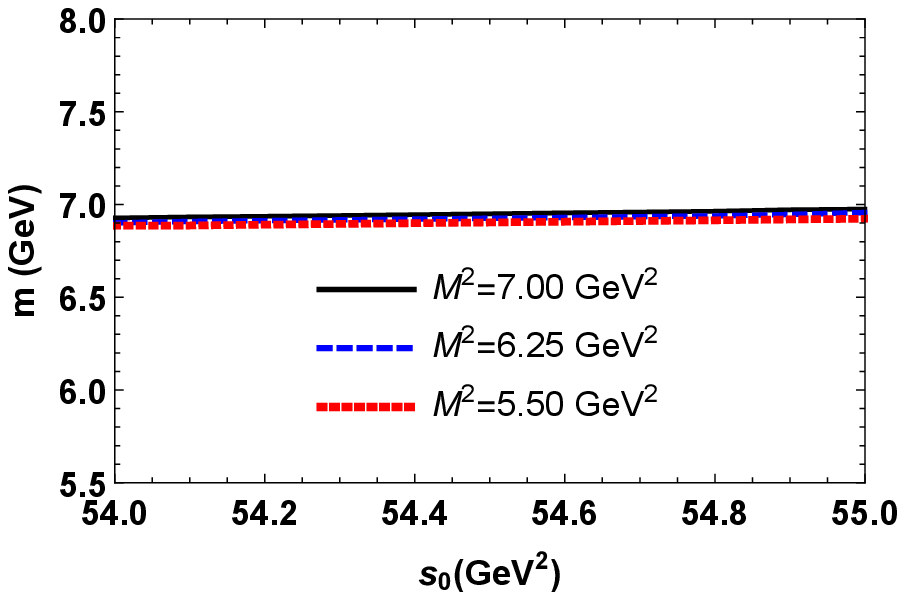}
\end{center}
\caption{ Mass of the tetraquark $T_{\mathrm{4c}}$ as a function of the
Borel $M^2$ (left), and the continuum threshold $s_0$ parameters (right).}
\label{fig:Mass}
\end{figure}

\end{widetext}

The region for $s_{0}$ has to comply with constraints coming from the
dominance of $\mathrm{PC}$ and prevalence of perturbative term in $\mathrm{%
OPE}$. Besides, $s_{0}$ separates contributions of the ground-level $T_{%
\mathrm{4c}}$ and first excited tetraquark $T_{\mathrm{4c}}^{\ast }$ with
the mass $m^{\ast }$, therefore the inequalities $m<\sqrt{s_{0}}<m^{\ast }$
should be satisfied. These restrictions provide an opportunity to verify
self-consistency of performed studies and correctness of the chosen region
for $s_{0}$. Additionally, using $\sqrt{s_{0}}<m^{\ast }$ it is possible to
estimate lower limit for the mass of the excited state $T_{\mathrm{4c}%
}^{\ast }$. In the lack of experimental and theoretical information on
parameters of excited tetraquarks, this is one of useful ways to gain some
knowledge about the mass of the first excited diquark-antidiquark state $T_{%
\mathrm{4c}}^{\ast }$.

The mass $m$ and current coupling $f$ of the tetraquark $T_{\mathrm{4c}}$
are calculated as values of these parameters averaged over the regions (\ref%
{eq:Wind1}). Our results for $m$ and $f$ are:
\begin{eqnarray}
m &=&(6928\pm 50)~\mathrm{MeV},  \notag \\
f &=&(2.06\pm 0.14)\times 10^{-2}~\mathrm{GeV}^{4}.  \label{eq:Result1}
\end{eqnarray}%
It is seen, that ambiguities of calculations are very small and are equal to
$\pm 1\%$ for the mass and $\pm 7\%$ in the case of the current coupling.
The results Eq.\ (\ref{eq:Result1}) correspond to sum rules' predictions at
the point $M^{2}=5.9~\mathrm{GeV}^{2}$ and $s_{0}=54.5~\mathrm{GeV}^{2}$,
where the pole contribution is $\mathrm{PC}\approx 0.7$ (see, Fig.\ \ref%
{fig:PC}). This fact ensures dominance of $\mathrm{PC}$ in extracted
quantities, and ground-state nature of the exotic meson $T_{\mathrm{4c}}$.
The mass $m$ of the tetraquark $T_{\mathrm{4c}}$ is plotted in Fig.\ \ref%
{fig:Mass}.

The dependence of the mass $m$ on the continuum threshold parameter $s_{0}$
can be seen in the right panel of Fig.\ \ref{fig:Mass}. It is not difficult
to be convinced in self-consistency of the present calculations. Indeed,
with the result for $m$ at hand, one sees that $m<\sqrt{s_{0}}$. It is also
easy to find $m^{\ast }>m+450~\mathrm{MeV}$, which for a particle containing
four $c$ quarks seem is a reasonable estimate.

Our result for $m$ is in excellent agreement with the mass of the resonance $%
X(6900)$ measured by the CMS collaboration. It is also compatible with LHCb
and ATLAS data though slightly exceeds them. But for detailed comparison
with available data, and more reliable conclusions on nature of $T_{\mathrm{%
4c}}$, we have to estimate its full width.

\begin{widetext}

\begin{figure}[h!]
\begin{center}
\includegraphics[totalheight=6cm,width=8cm]{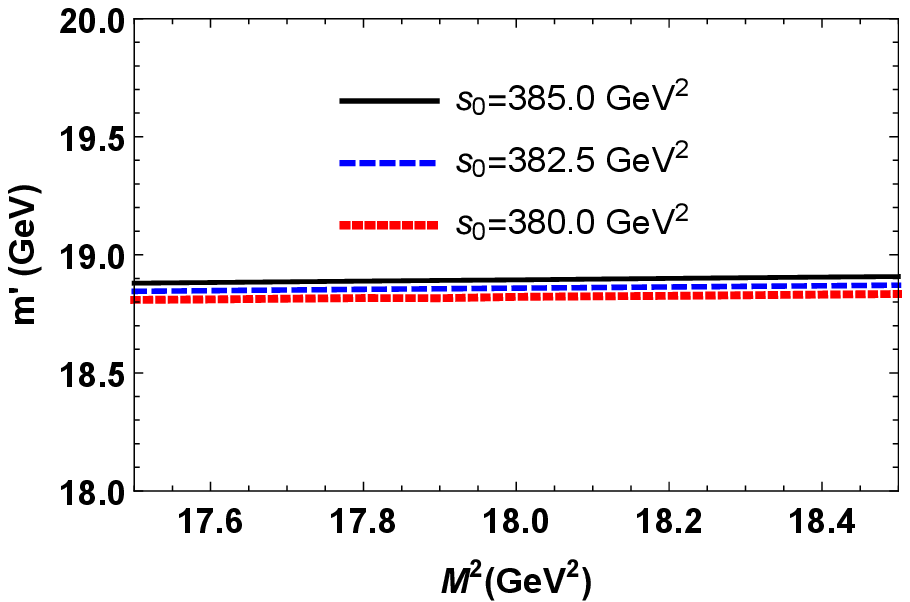}\,\, %
\includegraphics[totalheight=6cm,width=8cm]{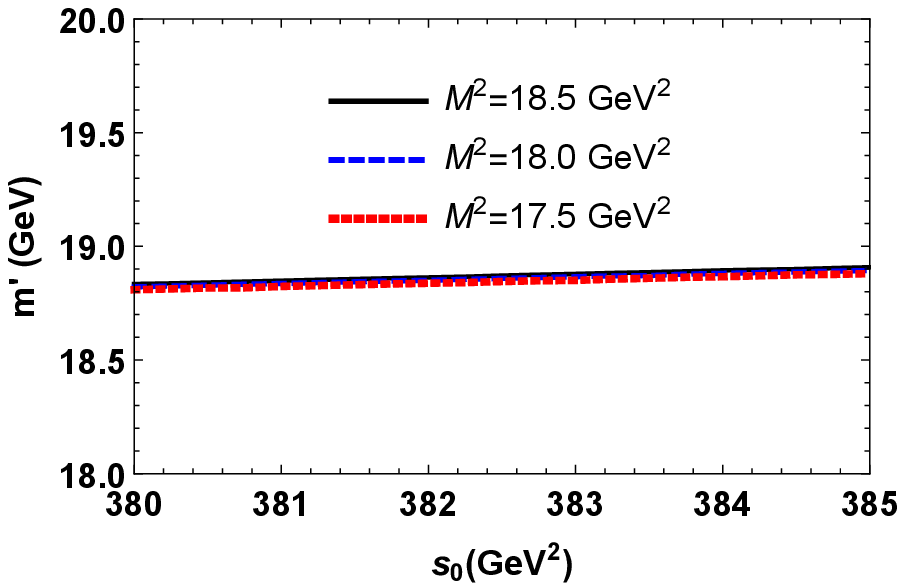}
\end{center}
\caption{ Mass $m^{\prime}$ of the tetraquark $T_{\mathrm{4b}}$ as a function of the parameters $M^2$ (left), and $s_0$ (right).}
\label{fig:MassB}
\end{figure}

\end{widetext}

In the case of the tetraquark $T_{\mathrm{4b}}$ for $M^{2}$ and $s_{0}$
computations give the following regions
\begin{eqnarray}
M^{2} &\in &[17.5,18.5]~\mathrm{GeV}^{2},  \notag \\
s_{0} &\in &[380,385]~\mathrm{GeV}^{2}.  \label{eq:Wind2}
\end{eqnarray}%
Here, the $\mathrm{PC}$ varies within the interval
\begin{equation}
0.64\geq \mathrm{PC}\geq 0.58.
\end{equation}%
The dimension-$4$ term constitutes $-2.2\%$ of the result at $M^{2}=17.5~%
\mathrm{GeV}^{2}$. The mass and current coupling of the fully beauty
tetraquark $T_{\mathrm{4b}}$ are equal to
\begin{eqnarray}
m^{\prime } &=&(18858\pm 50)~\mathrm{MeV},  \notag \\
f^{\prime } &=&(9.54\pm 0.71)\times 10^{-2}~\mathrm{GeV}^{4}.
\label{eq:Result2}
\end{eqnarray}%
Behavior of $m^{\prime }$ as a function of $M^{2}$ and $s_{0}$ is shown in
Fig.\ \ref{fig:MassB}.

The diquark-antidiquark state $T_{\mathrm{4b}}$ with $C\otimes C$ structure
and mass $m^{\prime }=18858~\mathrm{MeV}$ is above $\eta _{b}\eta _{b}$ but
below $\Upsilon (1S)\Upsilon (1S)$ thresholds. Consequently, it decays to a $%
\eta _{b}\eta _{b}$ pair and may be searched for in invariant mass
distributions of these mesons.

\section{Decays $T_{\mathrm{4c}}\rightarrow J/\protect\psi J/\protect\psi $
\ and $T_{\mathrm{4c}}\rightarrow J/\protect\psi \protect\psi ^{\prime }$}

\label{sec:Decays1}


The prediction for the mass of the diquark-antidiquark system with structure
$C\otimes C$ permits us to determine kinematically allowed decay modes of $%
T_{\mathrm{4c}}$. The mass $m=6928~\mathrm{MeV}$ of this tetraquark exceeds
the two-meson $J/\psi J/\psi $ and $J/\psi \psi ^{\prime }$ thresholds. The $%
m$ satisfies also the kinematical restrictions for productions of $\eta
_{c}\eta _{c}$ and $\eta _{c}\eta _{c}(2S)$ pairs. $T_{\mathrm{4c}}$ can
decay to conventional $\eta _{c}\chi _{c1}(1P)$ and $\chi _{c0}\chi _{c0}$
mesons as well. The decay $T_{\mathrm{4c}}\rightarrow \eta _{c}\chi
_{c1}(1P) $ is $P$-wave, whereas remaining five channels are $S$-wave
processes.

We start our studies from investigation of the decays $J/\psi J/\psi $ and $%
J/\psi \psi ^{\prime }$. The three-point sum rules for the strong form
factors $g_{1}(q^{2})$ and $g_{1}^{\ast }(q^{2})$ which describe interaction
of particles at vertices $T_{\mathrm{4c}}J/\psi J/\psi $ \ and $T_{\mathrm{4c%
}}J/\psi \psi ^{\prime }$ respectively, can be extracted from analysis of
the correlation function
\begin{eqnarray}
&&\Pi _{\mu \nu }(p,p^{\prime })=i^{2}\int d^{4}xd^{4}ye^{ip^{\prime
}y}e^{-ipx}\langle 0|\mathcal{T}\{J_{\mu }^{\psi }(y)  \notag \\
&&\times J_{\nu }^{\psi }(0)J^{\dagger }(x)\}|0\rangle ,  \label{eq:CF2}
\end{eqnarray}%
where the current $J(x)$ is defined by Eq.\ (\ref{eq:CR1}), and $J_{\mu
}^{\psi }(x)\ $is the interpolating currents for the mesons $J/\psi $ and $%
\psi ^{\prime }$
\begin{equation}
J_{\mu }^{\psi }(x)=\overline{c}_{i}(x)\gamma _{\mu }c_{i}(x),
\label{eq:CR2}
\end{equation}%
where $i=1,2,3$ are the color indices.

We apply usual recipes of the sum rule method and write down the correlation
function $\Pi _{\mu \nu }(p,p^{\prime })$ in terms of physical parameters of
particles. Because the tetraquark $T_{\mathrm{4c}}$ can decay both to $%
J/\psi J/\psi $ and $J/\psi \psi ^{\prime }$ pairs, we isolate in $\Pi _{\mu
\nu }(p,p^{\prime })$ contributions of the particles $J/\psi $ and $\psi
^{\prime }$ from higher resonances and continuum states' effects. Then, for
the phenomenological component of the sum rule $\Pi _{\mu \nu }^{\mathrm{Phys%
}}(p,p^{\prime })$, we get%
\begin{eqnarray}
&&\Pi _{\mu \nu }^{\mathrm{Phys}}(p,p^{\prime })=\frac{\langle 0|J_{\mu
}^{\psi }|J/\psi (p^{\prime })\rangle }{p^{\prime 2}-m_{1}^{2}}\frac{\langle
0|J_{\nu }^{\psi }|J/\psi (q)\rangle }{q^{2}-m_{1}^{2}}  \notag \\
&&\times \langle J/\psi (p^{\prime })J/\psi (q)|T_{\mathrm{4c}}(p)\rangle
\frac{\langle T_{\mathrm{4c}}(p)|J^{\dagger }|0\rangle }{p^{2}-m^{2}}  \notag
\\
&&+\frac{\langle 0|J_{\mu }^{\psi }|\psi (p^{\prime })\rangle }{p^{\prime
2}-m_{1}^{\ast 2}}\frac{\langle 0|J_{\nu }^{\psi }|J/\psi (q)\rangle }{%
q^{2}-m_{1}^{2}}  \notag \\
&&\times \langle \psi (p^{\prime })J/\psi (q)|T_{\mathrm{4c}}(p)\rangle
\frac{\langle T_{\mathrm{4c}}(p)|J^{\dagger }|0\rangle }{p^{2}-m^{2}}+\cdots
,  \label{eq:CF3}
\end{eqnarray}%
with $m_{1}$ and $m_{1}^{\ast }$ being the masses of $J/\psi $ and $\psi
^{\prime }$ mesons.

The function $\Pi _{\mu \nu }^{\mathrm{Phys}}(p,p^{\prime })$ can be
rewritten using the matrix elements of the tetraquark $T_{\mathrm{4c}}$, and
mesons $J/\psi $ and $\psi ^{\prime }$. The matrix element of $T_{\mathrm{4c}%
}$ is given by Eq.\ (\ref{eq:ME1}), whereas for $\langle 0|J_{\mu }^{\psi
}|J/\psi (p)\rangle $ and $\langle 0|J_{\mu }^{\psi }|\psi ^{\prime
}(p)\rangle $, we utilize
\begin{eqnarray}
\langle 0|J_{\mu }^{\psi }|J/\psi (p)\rangle &=&f_{1}m_{1}\varepsilon _{\mu
}(p),  \notag \\
\langle 0|J_{\mu }^{\psi }|\psi ^{\prime }(p)\rangle &=&f_{1}^{\ast
}m_{1}^{\ast }\widetilde{\varepsilon }_{\mu }(p),  \label{eq:ME2}
\end{eqnarray}%
where $f_{1}$, $f_{1}^{\ast }$, and $\varepsilon _{\mu }$, $\widetilde{%
\varepsilon }$ are the decay constants and polarization vectors of $J/\psi $
and $\psi ^{\prime }$, respectively.

For the vertices, we employ the following expressions%
\begin{eqnarray}
&&\langle J/\psi (p^{\prime })J/\psi (q)|T_{\mathrm{4c}}(p)\rangle
=g_{1}(q^{2})\left[ q\cdot p^{\prime }\varepsilon ^{\ast }(p^{\prime })\cdot
\varepsilon ^{\ast }(q)\right.  \notag \\
&&\left. -q\cdot \varepsilon ^{\ast }(p^{\prime })p^{\prime }\cdot
\varepsilon ^{\ast }(q)\right],  \label{eq:ME3}
\end{eqnarray}%
and
\begin{eqnarray}
\langle \psi (p^{\prime })J/\psi (q)|T_{\mathrm{4c}}(p)\rangle
&=&g_{1}^{\ast }(q^{2})\left[ q\cdot p^{\prime }\widetilde{\varepsilon }%
^{\ast }(p^{\prime })\cdot \varepsilon ^{\ast }(q)\right.  \notag \\
&&\left. -q\cdot \widetilde{\varepsilon }^{\ast }(p^{\prime })p^{\prime
}\cdot \varepsilon ^{\ast }(q)\right].  \label{eq:ME3A}
\end{eqnarray}

Having utilized these matrix elements and carried out simple calculations,
for $\Pi _{\mu \nu }^{\mathrm{Phys}}(p,p^{\prime })$ we find
\begin{eqnarray}
&&\Pi _{\mu \nu }^{\mathrm{Phys}}(p,p^{\prime })=g_{1}(q^{2})\frac{%
fmf_{1}^{2}m_{1}^{2}}{\left( p^{2}-m^{2}\right) \left( p^{\prime
2}-m_{1}^{2}\right) (q^{2}-m_{1}^{2})}  \notag \\
&&\times \left[ \frac{1}{2}\left( m^{2}-m_{1}^{2}-q^{2}\right) g_{\mu \nu
}-q_{\mu }p_{\nu }^{\prime }\right] +  \notag \\
&&+g_{1}^{\ast }(q^{2})\frac{fmf_{1}m_{1}f_{1}^{\ast }m_{1}^{\ast }}{\left(
p^{2}-m^{2}\right) \left( p^{\prime 2}-m_{1}^{\ast 2}\right)
(q^{2}-m_{1}^{2})}  \notag \\
&&\times \left[ \frac{1}{2}\left( m^{2}-m_{1}^{\ast 2}-q^{2}\right) g_{\mu
\nu }-q_{\mu }p_{\nu }^{\prime }\right] +\cdots,  \label{eq:CorrF5}
\end{eqnarray}%
where contributions of higher resonances and continuum states are denoted by
the dots. The function $\Pi _{\mu \nu }^{\mathrm{Phys}}(p,p^{\prime })$
contains the Lorentz structures, which are proportional to $g_{\mu \nu }$
and $q_{\mu }p_{\nu }^{\prime }$. Both of them can be employed to determine
the sum rules for $g_{1}(q^{2})$ and $g_{1}^{\ast }(q^{2})$. We work with
the structures $\sim g_{\mu \nu }$ and denote the relevant invariant
amplitudes by $\Pi _{1}^{\mathrm{Phys}}(p^{2},p^{\prime 2},q^{2})$ and $\Pi
_{2}^{\mathrm{Phys}}(p^{2},p^{\prime 2},q^{2})$, respectively. Then, a total
amplitude $\Pi ^{\mathrm{Phys}}(p^{2},p^{\prime 2},q^{2})$ is equal to a sum
of the functions $\Pi _{1,2}^{\mathrm{Phys}}(p^{2},p^{\prime 2},q^{2})$. The
Borel transformations of $\ \Pi ^{\mathrm{Phys}}(p^{2},p^{\prime 2},q^{2})$
over $-p^{2}$ and $-p^{\prime 2}$ give
\begin{eqnarray}
&&\mathcal{B}\Pi ^{\mathrm{Phys}}(p^{2},p^{\prime
2},q^{2})=g_{1}(q^{2})fmf_{1}^{2}m_{1}^{2}  \notag \\
&&\times \frac{m^{2}-m_{1}^{2}-q^{2}}{2(q^{2}-m_{1}^{2})}%
e^{-m^{2}/M_{1}^{2}}e^{-m_{1}^{2}/M_{2}^{2}}  \notag \\
&&+g_{1}^{\ast }(q^{2})fmf_{1}m_{1}f_{1}^{\ast }m_{1}^{\ast }\frac{%
m^{2}-m_{1}^{\ast 2}-q^{2}}{2(q^{2}-m_{1}^{2})}  \notag \\
&&\times e^{-m^{2}/M_{1}^{2}}e^{-m_{1}^{\ast 2}/M_{2}^{2}}+\cdots .
\label{eq:CorrF5a}
\end{eqnarray}

The correlation function $\Pi _{\mu \nu }(p,p^{\prime })$ expressed using
the $c$-quark propagators
\begin{eqnarray}
&&\Pi _{\mu \nu }^{\mathrm{OPE}}(p,p^{\prime })=2i^{2}\int
d^{4}xd^{4}ye^{ip^{\prime }y}e^{-ipx}  \notag \\
&&\times \left\{ \mathrm{Tr}\left[ \gamma _{\mu }S_{c}^{ib}(y-x)\widetilde{S}%
_{c}^{ja}(-x){}\gamma _{\nu }\widetilde{S}_{c}^{bj}(x)S_{c}^{ai}(x-y)\right]
\right.  \notag \\
&&\left. +\mathrm{Tr}\left[ \gamma _{\mu }S_{c}^{ia}(y-x)\widetilde{S}%
_{c}^{jb}(-x){}\gamma _{\nu }\widetilde{S}_{c}^{bj}(x)S_{c}^{ai}(x-y)\right]
\right\} ,  \notag \\
&&  \label{eq:QCDside}
\end{eqnarray}
is the QCD side of the sum rule.

\begin{table}[tbp]
\begin{tabular}{|c|c|}
\hline\hline
Parameters & Values (in $\mathrm{MeV}$) \\ \hline\hline
$m_1[m_{J/\psi}]$ & $3096.900 \pm 0.006$ \\
$f_1[f_{J/\psi}]$ & $409 \pm 15$ \\
$m_{1}^{\ast}[m_{\psi^{\prime}}]$ & $3686.10 \pm 0.06$ \\
$f_{1}^{\ast}[f_{\psi^{\prime}}]$ & $279 \pm 8$ \\
$m_2[m_{\eta_c}]$ & $2983.9 \pm 0.4$ \\
$f_2[f_{\eta_c}]$ & $398.1 \pm 1.0$ \\
$m_{2}^{\ast}[m_{\eta_c(2S)}]$ & $3637.5 \pm 1.1$ \\
$f_{2}^{\ast}[f_{\eta_c(2S)}]$ & $331 $ \\
$m_3[m_{\chi _{c1}}]$ & $3510.67 \pm 0.05$ \\
$f_3[f_{\chi _{c1}}]$ & $344 \pm 27$ \\
$m_4[m_{\chi_{c0}}]$ & $3414.71 \pm 0.30$ \\
$f_4[f_{\chi_{c0}}]$ & $343 $ \\ \hline
$m_5[m_{\eta_b}]$ & $9398.7 \pm 2.0$ \\
$f_5[f_{\eta_b}]$ & $724 \pm 12$ \\ \hline\hline
\end{tabular}%
\caption{Masses and decay constants of the various charmonia and $\protect%
\eta _{b}$ meson, which have been used in numerical computations. }
\label{tab:Param}
\end{table}
The invariant amplitude $\Pi ^{\mathrm{OPE}}(p^{2},p^{\prime 2},q^{2})$
obtained from the component $\sim g_{\mu \nu }$ of Eq.\ (\ref{eq:QCDside})
can be used in following studies. By equating the double Borel transforms of
the amplitudes $\Pi ^{\mathrm{OPE}}(p^{2},p^{\prime 2},q^{2})$ and $\Pi ^{%
\mathrm{Phys}}(p^{2},p^{\prime 2},q^{2})$, and carrying out the continuum
subtraction, one can find the sum rules for the form factors $g_{1}(q^{2})$
and $g_{1}^{\ast }(q^{2})$.

After the Borel transformation and continuum subtraction, $\Pi ^{\mathrm{OPE}%
}(p^{2},p^{\prime 2},q^{2})$ can be written down in terms of the spectral
density $\rho (s,s^{\prime },q^{2})$ determined as a relevant imaginary part
of $\Pi _{\mu \nu }^{\mathrm{OPE}}(p,p^{\prime })$
\begin{eqnarray}
&&\Pi (\mathbf{M}^{2},\mathbf{s}_{0},q^{2})=\int_{16m_{c}^{2}}^{s_{0}}ds%
\int_{4m_{c}^{2}}^{s_{0}^{\prime }}ds^{\prime }\rho (s,s^{\prime },q^{2})
\notag \\
&&\times e^{-s/M_{1}^{2}}e^{-s^{\prime }/M_{2}^{2}}.  \label{eq:SCoupl}
\end{eqnarray}%
Here, $\mathbf{M}^{2}=(M_{1}^{2},M_{2}^{2})$ and $\mathbf{s}%
_{0}=(s_{0},s_{0}^{\prime })$ are the Borel and continuum threshold
parameters, respectively. The pair of parameters $(M_{1}^{2},s_{0})$
corresponds to the tetraquark channel, the pair ($M_{2}^{2}$, $s_{0}^{\prime
}$) -- to the $J/\psi $ ($\psi ^{\prime }$) channel.

To extract the sum rules for $g_{1}(q^{2})$ and $g_{1}^{\ast }(q^{2})$, at
the first stage of analysis, we fix the continuum subtraction parameter as $%
4m_{c}^{2}<s_{0}^{\prime }<m_{1}^{\ast 2}$, where $m_{1}^{\ast }$ is the
mass of the excited $\psi ^{\prime }=\psi (2^{3}S_{1})$ meson. By this way,
we include the second term in Eq.\ (\ref{eq:CorrF5a}) into higher resonances
and continuum states. This scheme is the standard "ground-state + continuum"
approach, when the physical side of the sum rule contains a contribution
coming only from ground-state particles. Then, it is not difficult to derive
the sum rule for the form factor $g_{1}(q^{2})$
\begin{eqnarray}
&&g_{1}(q^{2})=\frac{2}{fmf_{1}^{2}m_{1}^{2}}\frac{q^{2}-m_{1}^{2}}{%
m^{2}-m_{1}^{2}-q^{2}}  \notag \\
&&\times e^{m^{2}/M_{1}^{2}}e^{m_{1}^{2}/M_{2}^{2}}\Pi (\mathbf{M}^{2},%
\mathbf{s}_{0},q^{2}).  \label{eq:SRCoup}
\end{eqnarray}

At the next phase of calculations, we choose $m_{1}^{\ast 2}<s_{0}^{\ast
\prime }<m^{\ast \ast 2}$, with $m^{\ast \ast }=(4039\pm 1)~\mathrm{MeV}$
being the mass of the next excited state $\psi (3^{3}S_{1})$ \cite%
{PDG:2022,Barnes:2005pb}. By this way, we include into consideration the
second term in Eq.\ (\ref{eq:CorrF5a}): This is "ground-state + excited
state + continuum" scheme. Afterwards, by employing results obtained for $%
g_{1}(q^{2})$ one can determine $g_{1}^{\ast }(q^{2})$.

All information necessary for numerical computations of the form factors $%
g_{1}(q^{2})$ and $g_{1}^{\ast }(q^{2})$ are collected in Table\ \ref%
{tab:Param}. In this Table, we present parameters of the $\eta _{c}$, $\eta
_{c}(2S)$, $\chi _{c1}(1P)$, and $\eta _{b}$ mesons that will be used later
to explore decays of $T_{\mathrm{4c}}$ and $T_{\mathrm{4b}}$. The masses of
the particles are borrowed from Ref.\ \cite{PDG:2022}. For decay constant of
the meson $J/\psi $, we employ the experimental value reported in Ref.\ \cite%
{Kiselev:2001xa}. As $f_{\eta _{c}}$ and $f_{\eta _{b}}$, we use results of
QCD lattice simulations \cite{Hatton:2020qhk,Hatton:2021dvg}, whereas for $%
f_{\chi _{c1}}$ and $f_{\chi _{c0}}$-- the sum rule predictions from Refs.
\cite{VeliVeliev:2012cc,Veliev:2010gb}.

To perform numerical computations, we also have to choose the working
regions for the parameters $\mathbf{M}^{2}$ and $\mathbf{s}_{0}$. They
should meet standard constraints of sum rule calculations, which have been
discussed in Sec.\ \ref{sec:Masses}. For $M_{1}^{2}$ and $s_{0}$ which are
actual for the $T_{\mathrm{4c}}$ channel, we employ the working regions Eq.\
(\ref{eq:Wind1}). The parameters $(M_{2}^{2},\ s_{0}^{\prime })$ for the $%
J/\psi $ channel are changed within limits%
\begin{equation}
M_{2}^{2}\in \lbrack 4,5]~\mathrm{GeV}^{2},\ s_{0}^{\prime }\in \lbrack
12,13]~\mathrm{GeV}^{2}.  \label{eq:Wind3}
\end{equation}%
In the second phase of analysis, we use
\begin{equation}
M_{2}^{2}\in \lbrack 4,5]~\mathrm{GeV}^{2},\ s_{0}^{\ast \prime }\in \lbrack
15,16]~\mathrm{GeV}^{2}.
\end{equation}

\begin{figure}[h]
\includegraphics[width=8.5cm]{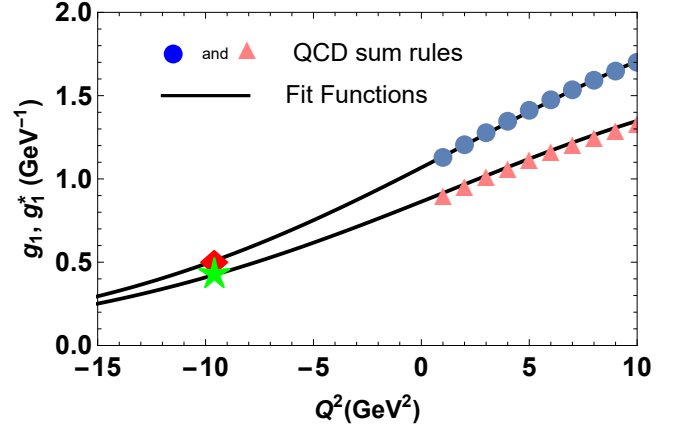}
\caption{The sum rule predictions and fit functions for the form factors $%
g_{1}(Q^{2})$ (upper line)and $g_{1}^{\ast}(Q^{2})$ (lower line). The red
diamond and green star denote the point $Q^{2}=-m_{1}^{2}$, where the strong
couplings $g_{1}$ and $g_{1}^{\ast}$ are evaluated. }
\label{fig:Fit}
\end{figure}

The sum rule method gives reliable predictions in the deep-Euclidean region $%
q^{2}<0$. Therefore, it is convenient to introduce a new variable $%
Q^{2}=-q^{2}$ and denote the obtained function by $g_{1}(Q^{2})$. The range
of $Q^{2}$ analyzed by the sum rule method covers the interval $Q^{2}=1-10~%
\mathrm{GeV}^{2}$. Results of calculations are plotted in Fig.\ \ref{fig:Fit}%
.

But the width of the decay $T_{\mathrm{4c}}\rightarrow J/\psi J/\psi $ is
determined by the form factor $g_{1}(q^{2})$ at the mass shell $%
q^{2}=m_{1}^{2}$. Stated differently, one has to find $%
g_{1}(Q^{2}=-m_{1}^{2})$. To avoid this problem, we use a fit function $%
\mathcal{G}_{1}(Q^{2})$, which at momenta $Q^{2}>0$ gives the same values as
the sum rule calculations, but can be extrapolated to the region of $Q^{2}<0$%
. To this end, we employ functions $\mathcal{G}_{i}^{(\ast )}(Q^{2})$
\begin{equation}
\mathcal{G}_{i}^{(\ast )}(Q^{2})=\mathcal{G}_{i}^{0(\ast )}\mathrm{\exp }%
\left[ c_{i}^{1(\ast )}\frac{Q^{2}}{m^{2}}+c_{i}^{2(\ast )}\left( \frac{Q^{2}%
}{m^{2}}\right) ^{2}\right],  \label{eq:FitF}
\end{equation}%
with parameters $\mathcal{G}_{i}^{0(\ast )}$, $c_{i}^{1(\ast )}$ and $%
c_{i}^{2(\ast )}$. Calculations demonstrate that $\mathcal{G}_{1}^{0}=1.07~%
\mathrm{GeV}^{-1}$, $c_{1}^{1}=2.99$, and $c_{1}^{2}=-3.57$ give a nice
agreement with the sum rules data for $g_{1}(Q^{2})$ shown in Fig.\ \ref%
{fig:Fit}.

At the mass shell $q^{2}=m_{1}^{2}$ the function $\mathcal{G}_{1}(Q^{2})$ is
equal to
\begin{equation}
g_{1}\equiv \mathcal{G}_{1}(-m_{1}^{2})=(5.1\pm 1.1)\times 10^{-1}\ \mathrm{%
GeV}^{-1}.  \label{eq:Coup1}
\end{equation}%
Partial width of the process $T_{\mathrm{4c}}\rightarrow J/\psi J/\psi $ can
be evaluated by employing the following expression
\begin{equation}
\Gamma \left[ T_{\mathrm{4c}}\rightarrow J/\psi J/\psi \right] =g_{1}^{2}%
\frac{\lambda _{1}}{8\pi }\left( \frac{m_{1}^{4}}{m^{2}}+\frac{2\lambda
_{1}^{2}}{3}\right),  \label{eq:PartDW}
\end{equation}%
where $\lambda _{1}=\lambda (m,m_{1},m_{1})$ and
\begin{equation}
\lambda (a,b,c)=\frac{\sqrt{%
a^{4}+b^{4}+c^{4}-2(a^{2}b^{2}+a^{2}c^{2}+b^{2}c^{2})}}{2a}.
\end{equation}%
Then it is easy to find that
\begin{equation}
\Gamma \left[ T_{\mathrm{4c}}\rightarrow J/\psi J/\psi \right] =(56\pm 18)~%
\mathrm{MeV}.  \label{eq:DW1}
\end{equation}

The decay $T_{\mathrm{4c}}\rightarrow J/\psi \psi ^{\prime }$ can be
explored in accordance with a scheme described above. In this case, the
extrapolating function $\mathcal{G}_{1}^{\ast }(Q^{2})$ has the parameters $%
\mathcal{G}_{1}^{0\ast }=0.86~\mathrm{GeV}^{-1}$, $c_{1}^{1\ast }=2.86$, and
$c_{1}^{2\ast }=-3.47$. The sum rule predictions for the form factor $%
g_{1}^{\ast }(q^{2})$, as well as the function $\mathcal{G}_{1}^{\ast
}(Q^{2})$ are depicted in Fig.\ \ref{fig:Fit}. One can be convinced in a
reasonable agreement between the sum rules data and $\mathcal{G}_{1}^{\ast
}(Q^{2})$.

The strong coupling $g_{1}^{\ast }$ is calculated at the mass shell $%
q^{2}=m_{1}^{2}$
\begin{equation}
g_{1}^{\ast }\equiv \mathcal{G}_{1}^{\ast }(-m_{1}^{2})=(4.2\pm 1.0)\times
10^{-1}\ \mathrm{GeV}^{-1}.  \label{eq:Coup2}
\end{equation}%
The width of the process is found by means of Eq.\ (\ref{eq:PartDW}) with
the following substitutions $g_{1}\rightarrow g_{1}^{\ast }$, $\lambda
_{1}\rightarrow \lambda _{2}=\lambda (m,m_{1}^{\ast },m_{1})$, and $%
m_{1}^{4}\rightarrow m_{1}^{2}m_{1}^{\ast 2}$ which leads to the result
\begin{equation}
\Gamma \left[ T_{\mathrm{4c}}\rightarrow J/\psi \psi ^{\prime }\right]
=(15\pm 5)~\mathrm{MeV}.  \label{eq:DW1A}
\end{equation}%
Essential parameters of these decays are shown in Table \ref{tab:Channels}.


\section{Processes $T_{\mathrm{4c}}\rightarrow \protect\eta _{c}\protect\eta %
_{c}$, $\protect\eta _{c}\protect\eta _{c}(2S)$}

\label{sec:Decays2}

The decays $T_{\mathrm{4c}}\rightarrow \eta _{c}\eta _{c}$ and $T_{\mathrm{4c%
}}\rightarrow \eta _{c}\eta _{c}(2S)$ can be explored in a similar way. The
strong couplings $g_{2}$ and $g_{2}^{\ast }$ which correspond to the
vertices $T_{\mathrm{4c}}\eta _{c}\eta _{c}$ and $T_{\mathrm{4c}}\eta
_{c}\eta _{c}(2S)$ can be extracted from the correlation function
\begin{eqnarray}
&&\Pi (p,p^{\prime })=i^{2}\int d^{4}xd^{4}ye^{ip^{\prime }y}e^{-ipx}\langle
0|\mathcal{T}\{J^{\eta _{c}}(y)  \notag \\
&&\times J^{\eta _{c}}(0)J^{\dagger }(x)\}|0\rangle,  \label{eq:CF4}
\end{eqnarray}%
where the current $J^{\eta _{c}}(x)$ is
\begin{equation}
J^{\eta _{c}}(x)=\overline{c}_{i}(x)i\gamma _{5}c_{i}(x).  \label{eq:CR3}
\end{equation}%
Separating the ground-level and first excited state contributions from
effects of higher resonances and continuum states, we write the correlation
function (\ref{eq:CF4}) in the following form%
\begin{eqnarray}
&&\Pi ^{\mathrm{Phys}}(p,p^{\prime })=\frac{\langle 0|J^{\eta _{c}}|\eta
_{c}(p^{\prime })\rangle }{p^{\prime 2}-m_{2}^{2}}\frac{\langle 0|J^{\eta
_{c}}|\eta _{c}(q)\rangle }{q^{2}-m_{2}^{2}}  \notag \\
&&\times \langle \eta _{c}(p^{\prime })\eta _{c}(q)|T_{\mathrm{4c}%
}(p)\rangle \frac{\langle T_{\mathrm{4c}}(p)|J^{\dagger }|0\rangle }{%
p^{2}-m^{2}}  \notag \\
&&+\frac{\langle 0|J^{\eta _{c}}|\eta _{c}(2S)(p^{\prime })\rangle }{%
p^{\prime 2}-m_{2}^{\ast 2}}\frac{\langle 0|J^{\eta _{c}}|\eta
_{c}(q)\rangle }{q^{2}-m_{2}^{2}}  \notag \\
&&\times \langle \eta _{c}(2S)(p^{\prime })\eta _{c}(q)|T_{\mathrm{4c}%
}(p)\rangle \frac{\langle T_{\mathrm{4c}}(p)|J^{\dagger }|0\rangle }{%
p^{2}-m^{2}}+\cdots,  \label{eq:CF5}
\end{eqnarray}%
where $m_{2}$ and $m_{2}^{\ast }$ are the masses of the $\eta _{c}$ and $%
\eta _{c}(2S)$ mesons. The matrix elements of scalar and two pseudoscalar
particles'\ vertices are modeled in the form
\begin{eqnarray}
&&\langle \eta _{c}(p^{\prime })\eta _{c}(q)|T_{\mathrm{4c}}(p)\rangle
=g_{2}(q^{2})p\cdot p^{\prime },  \notag \\
&&\langle \eta _{c}(2S)(p^{\prime })\eta _{c}(q)|T_{\mathrm{4c}}(p)\rangle
=g_{2}^{\ast }(q^{2})p\cdot p^{\prime }.  \label{eq:ME5}
\end{eqnarray}%
To express the correlator $\Pi ^{\mathrm{Phys}}(p,p^{\prime })$ in terms of
physical parameters of the particles $T_{\mathrm{4c}}$, $\eta _{c}$, and $%
\eta _{c}(2S)$, we use the matrix element Eq.\ (\ref{eq:ME1}) and
\begin{equation}
\langle 0|J^{\eta _{c}}|\eta _{c}\rangle =\frac{f_{2}m_{2}^{2}}{2m_{c}}%
,\langle 0|J^{\eta _{c}}|\eta _{c}(2S)\rangle =\frac{f_{2}^{\ast
}m_{2}^{\ast 2}}{2m_{c}},  \label{eq:ME4}
\end{equation}%
with $f_{2}$ and $f_{2}^{\ast }$ being the decay constants of the mesons $%
\eta _{c}$ and $\eta _{c}(2S)$, respectively. The correlation function $\Pi
^{\mathrm{Phys}}(p,p^{\prime })$ then takes the form%
\begin{eqnarray}
&&\Pi ^{\mathrm{Phys}}(p,p^{\prime })=\frac{g_{2}(q^{2})fmf_{2}^{2}m_{2}^{4}%
}{8m_{c}^{2}\left( p^{2}-m^{2}\right) \left( p^{\prime 2}-m_{2}^{2}\right) }
\notag \\
&&\times \frac{m^{2}+m_{2}^{2}-q^{2}}{q^{2}-m_{2}^{2}}+\frac{g_{2}^{\ast
}(q^{2})fmf_{2}m_{2}^{2}f_{2}^{\ast }m_{2}^{\ast 2}}{8m_{c}^{2}\left(
p^{2}-m^{2}\right) \left( p^{\prime 2}-m_{2}^{\ast 2}\right) }  \notag \\
&&\times \frac{m^{2}+m_{2}^{\ast 2}-q^{2}}{q^{2}-m_{2}^{2}}+\cdots .
\label{eq:CF6}
\end{eqnarray}%
The correlation function $\Pi ^{\mathrm{Phys}}(p,p^{\prime })$ has simple
Lorentz structure proportional to $\mathrm{I}$, hence right-hand side of
Eq.\ (\ref{eq:CF6}) is the corresponding invariant amplitude $\widetilde{\Pi
}^{\mathrm{Phys}}(p^{2},p^{\prime 2},q^{2})$.

Using quark-gluon degrees of freedom, we can find the QCD side of the sum
rule
\begin{eqnarray}
&&\Pi ^{\mathrm{OPE}}(p,p^{\prime })=2i^{2}\int d^{4}xd^{4}ye^{ip^{\prime
}y}e^{-ipx}  \notag \\
&&\left\{ \times \mathrm{Tr}\left[ \gamma _{5}S_{c}^{ia}(y-x)\widetilde{S}%
_{c}^{jb}(-x){}\gamma _{5}\widetilde{S}_{c}^{bj}(x)S_{c}^{ai}(x-y)\right]
\right.  \notag \\
&&\left. +\mathrm{Tr}\left[ \gamma _{5}S_{c}^{ib}(y-x)\widetilde{S}%
_{c}^{ja}(-x){}\gamma _{5}\widetilde{S}_{c}^{bj}(x)S_{c}^{ai}(x-y)\right]
\right\} .  \notag \\
&&  \label{eq:QCDside2}
\end{eqnarray}%
The sum rule for the strong form factor $g_{2}(q^{2})$ reads%
\begin{eqnarray}
&&g_{2}(q^{2})=\frac{8m_{c}^{2}}{fmf_{2}^{2}m_{2}^{4}}\frac{q^{2}-m_{2}^{2}}{%
m^{2}+m_{2}^{2}-q^{2}}  \notag \\
&&\times e^{m^{2}/M_{1}^{2}}e^{m_{2}^{2}/M_{2}^{2}}\widetilde{\Pi }(\mathbf{M%
}^{2},\mathbf{s}_{0},q^{2}),  \label{eq:SRCoup2}
\end{eqnarray}%
with $\widetilde{\Pi }(\mathbf{M}^{2},\mathbf{s}_{0},q^{2})$ being the
invariant amplitude $\widetilde{\Pi }^{\mathrm{OPE}}(p^{2},p^{\prime
2},q^{2})$ corresponding to the correlator $\Pi ^{\mathrm{OPE}}(p,p^{\prime
})$ after the Borel transformations and subtractions. Numerical computations
are carried out using Eq.\ (\ref{eq:SRCoup2}), parameters of the meson $\eta
_{c}$ from Table\ \ref{tab:Param}, and working regions for $\mathbf{M}^{2}$
and $\mathbf{s}_{0}$. The Borel and continuum subtraction parameters $%
M_{1}^{2}$ and $s_{0}$ in the $T_{\mathrm{4c}}$ channel are chosen as in
Eq.\ (\ref{eq:Wind1}), whereas for $M_{2}^{2}$ and $s_{0}^{\prime }$ which
correspond to the $\eta _{c}$ channel, we employ
\begin{equation}
M_{2}^{2}\in \lbrack 3.5,4.5]~\mathrm{GeV}^{2},\ s_{0}^{\prime }\in \lbrack
11,12]~\mathrm{GeV}^{2}.  \label{eq:Wind4}
\end{equation}

The interpolating function $\mathcal{G}_{2}(Q^{2})$ has the following
parameters: $\mathcal{G}_{2}^{0}=0.38~\mathrm{GeV}^{-1}$, $c_{2}^{1}=3.62$,
and $c_{2}^{2}=-4.17$. For the strong coupling $g_{2}$, we get
\begin{equation}
g_{2}\equiv \mathcal{G}_{2}(-m_{2}^{2})=(1.7\pm 0.4)\times 10^{-1}\ \mathrm{%
GeV}^{-1}.
\end{equation}%
The width of the process $T_{\mathrm{4c}}\rightarrow \eta _{c}\eta _{c}$ is
determined by means of the formula%
\begin{equation}
\Gamma \left[ T_{\mathrm{4c}}\rightarrow \eta _{c}\eta _{c}\right] =g_{2}^{2}%
\frac{m_{2}^{2}\lambda _{2}}{8\pi }\left( 1+\frac{\lambda _{2}^{2}}{m_{2}^{2}%
}\right),  \label{eq:PDw2}
\end{equation}%
where $\lambda _{2}=\lambda (m,m_{2},m_{2})$. Finally, we obtain
\begin{equation}
\Gamma \left[ T_{\mathrm{4c}}\rightarrow \eta _{c}\eta _{c}\right] =(24\pm
8)~\mathrm{MeV}.  \label{eq:DW2}
\end{equation}%
For the channel $T_{\mathrm{4c}}\rightarrow \eta _{c}\eta _{c}(2S)$, we use
\begin{equation}
M_{2}^{2}\in \lbrack 3.5,4.5]~\mathrm{GeV}^{2},\ s_{0}^{\ast \prime }\in
\lbrack 13,14]~\mathrm{GeV}^{2},  \label{eq:Wind4A}
\end{equation}%
and find
\begin{equation}
g_{2}^{\ast }\equiv \mathcal{G}_{2}^{\ast }(-m_{2}^{2})=(9.0\pm 2.8)\times
10^{-2}\ \mathrm{GeV}^{-1}.
\end{equation}%
The $g_{2}^{\ast }$ is evaluated using the fit function $\mathcal{G}%
_{2}^{\ast }(Q^{2})$ with the parameters $\mathcal{G}_{2}^{0\ast }=0.21~%
\mathrm{GeV}^{-1}$, $c_{2}^{1\ast }=3.62$, and $c_{2}^{2\ast }=-4.17$. The
width of this decay is equal to
\begin{equation}
\Gamma \left[ T_{\mathrm{4c}}\rightarrow \eta _{c}\eta _{c}(2S)\right]
=(5\pm 2)~\mathrm{MeV}.
\end{equation}


\section{Decays $T_{\mathrm{4c}}\rightarrow \protect\eta _{c}\protect\chi %
_{c1}(1P)$ and $T_{\mathrm{4c}}\rightarrow \protect\chi _{c0}\protect\chi %
_{c0}$}

\label{sec:Decays2A}

In this section, we study the processes $T_{\mathrm{4c}}\rightarrow \eta
_{c}\chi _{c1}(1P)$ and $T_{\mathrm{4c}}\rightarrow \chi _{c0}\chi _{c0}$,
which are the $P$- and $S$-wave decays of the tetraquark $T_{\mathrm{4c}}$,
respectively. The two-meson thresholds for these decay channels are equal to
$6495~\mathrm{MeV}$ and $6830~\mathrm{MeV}$, hence they are kinematically
allowed modes of $T_{\mathrm{4c}}$.


\subsection{$T_{\mathrm{4c}}\rightarrow \protect\eta _{c}\protect\chi %
_{c1}(P)$}


Treatment of the $P$-wave process $T_{\mathrm{4c}}\rightarrow \eta _{c}\chi
_{c1}(P)$ is performed in the context of the standard method. The
three-point correlator to be considered in this case is
\begin{eqnarray}
&&\Pi _{\mu }(p,p^{\prime })=i^{2}\int d^{4}xd^{4}ye^{ip^{\prime
}y}e^{-ipx}\langle 0|\mathcal{T}\{J_{\mu }^{\chi _{c1}}(y)  \notag \\
&&\times J^{\eta _{c}}(0)J^{\dagger }(x)\}|0\rangle,  \label{eq:CF7}
\end{eqnarray}%
where $J_{\mu }^{\chi _{c1}}(y)$ is the interpolating current for the meson $%
\chi _{c1}(1P)$%
\begin{equation}
J_{\mu }^{\chi _{c1}}(y)=\overline{c}_{j}(x)\gamma _{5}\gamma _{\mu
}c_{j}(x).  \label{eq:Curr1}
\end{equation}

In terms of the physical parameters of involved particles the correlation
function has the form%
\begin{eqnarray}
&&\Pi _{\mu }^{\mathrm{Phys}}(p,p^{\prime })=g_{3}(q^{2})\frac{%
fmf_{2}m_{2}^{2}f_{3}m_{3}}{2m_{c}\left( p^{2}-m^{2}\right) \left( p^{\prime
2}-m_{3}^{2}\right) }  \notag \\
&&\times \frac{1}{q^{2}-m_{2}^{2}}\left[ \frac{m^{2}-m_{3}^{2}-q^{2}}{%
2m_{3}^{2}}p_{\mu }^{\prime }-q_{\mu }\right] +\cdots.  \label{eq:CF8}
\end{eqnarray}%
In Eq.\ (\ref{eq:CF8}) $m_{3}$ and $f_{3}$ are the mass and decay constant
of the meson $\chi _{c1}(1P)$, respectively. To derive the correlator $\Pi
_{\mu }^{\mathrm{Phys}}(p,p^{\prime })$, we have used the known matrix
elements of the tetraquark $T_{\mathrm{4c}}$ and meson $\eta _{c}$, as well
as new matrix elements
\begin{equation}
\langle 0|J_{\mu }^{\chi _{c1}}|\chi _{c1}(p^{\prime })\rangle
=f_{3}m_{3}\varepsilon _{\mu }^{\ast }(p^{\prime }),
\end{equation}%
and
\begin{equation}
\langle \eta _{c}(q)\chi _{c1}(p^{\prime })|T_{\mathrm{4c}}(p)\rangle
=g_{3}(q^{2})p\cdot \varepsilon ^{\ast }(p^{\prime }),
\end{equation}%
where $\varepsilon _{\mu }^{\ast }(p^{\prime })$ is the polarization vector
of $\chi _{c1}(1P)$.

The QCD side $\Pi _{\mu }^{\mathrm{OPE}}(p,p^{\prime })$ is given by the
formula
\begin{eqnarray}
&&\Pi _{\mu }^{\mathrm{OPE}}(p,p^{\prime })=2i\int d^{4}xd^{4}ye^{ip^{\prime
}y}e^{-ipx}  \notag \\
&&\times \left\{ \mathrm{Tr}\left[ \gamma _{\mu }\gamma _{5}S_{c}^{ia}(y-x)%
\widetilde{S}_{c}^{jb}(-x){}\gamma _{5}\widetilde{S}%
_{c}^{bj}(x)S_{c}^{ai}(x-y)\right] \right.  \notag \\
&&\left. +\mathrm{Tr}\left[ \gamma _{\mu }\gamma _{5}S_{c}^{ib}(y-x)%
\widetilde{S}_{c}^{ja}(-x){}\gamma _{5}\widetilde{S}%
_{c}^{bj}(x)S_{c}^{ai}(x-y)\right] \right\} .  \notag \\
&&  \label{eq:CF9}
\end{eqnarray}%
The sum rule for $g_{3}(q^{2})$ is derived using invariant amplitudes
corresponding to terms $\sim p_{\mu }^{\prime }$ in $\Pi _{\mu }^{\mathrm{%
Phys}}(p,p^{\prime })$ and $\Pi _{\mu }^{\mathrm{OPE}}(p,p^{\prime })$.

In numerical analysis, the parameters $M_{2}^{2}$ and $s_{0}^{\prime }$ in
the $\chi _{c1}$ channel are chosen in the following way
\begin{equation}
M_{2}^{2}\in \lbrack 4,5]~\mathrm{GeV}^{2},\ s_{0}^{\prime }\in \lbrack
13,14]~\mathrm{GeV}^{2}.  \label{eq:Wind5}
\end{equation}%
For the parameters of the fit function $\mathcal{G}_{3}(Q^{2})$, we get $%
\mathcal{G}_{3}^{0}=5.16$, $c_{3}^{1}=3.16$, and $c_{3}^{2}=-3.87$. Then,
the strong coupling $g_{3}$ is equal to
\begin{equation}
g_{3}\equiv \mathcal{G}_{3}(-m_{2}^{2})=2.5\pm 0.6.
\end{equation}%
The width of the decay $T_{\mathrm{4c}}\rightarrow \eta _{c}\chi _{c1}(P)$ \
can be calculated by mean of the expression
\begin{equation}
\Gamma \left[ T_{\mathrm{4c}}\rightarrow \eta _{c}\chi _{c1}(P)\right]
=g_{3}^{2}\frac{\lambda _{3}^{3}}{24\pi m_{3}^{2}},  \label{eq:DW3}
\end{equation}%
where $\lambda _{3}=\lambda (m,m_{3},m_{2})$. The width of this process is
equal to
\begin{equation}
\Gamma \left[ T_{\mathrm{4c}}\rightarrow \eta _{c}\chi _{c1}(P)\right]
=(12\pm 4)~\mathrm{MeV}.  \label{eq:DW4}
\end{equation}


\subsection{$T_{\mathrm{4c}}\rightarrow \protect\chi _{c0}\protect\chi _{c0}$%
}


To explore the decay $T_{\mathrm{4c}}\rightarrow \chi _{c0}\chi _{c0}$, we
consider the correlation function
\begin{eqnarray}
&&\Pi (p,p^{\prime })=i^{2}\int d^{4}xd^{4}ye^{ip^{\prime }y}e^{-ipx}\langle
0|\mathcal{T}\{J^{\chi _{c0}}(y)  \notag \\
&&\times J^{\chi _{c0}}(0)J^{\dagger }(x)\}|0\rangle,
\end{eqnarray}%
with $J^{\chi _{c0}}(x)$ being the interpolating current for the scalar
meson $\chi _{c0}$%
\begin{equation}
J^{\chi _{c0}}(x)=\overline{c}_{i}(x)c_{i}(x).
\end{equation}

The physical side $\Pi ^{\mathrm{Phys}}(p,p^{\prime })$ of the sum rule is
\begin{eqnarray}
&&\Pi ^{\mathrm{Phys}}(p,p^{\prime })=g_{4}(q^{2})\frac{fmf_{4}m_{4}}{%
2\left( p^{2}-m^{2}\right) \left( p^{\prime 2}-m_{4}^{2}\right) }  \notag \\
&&\times \frac{m^{2}+m_{4}^{2}-q^{2}}{q^{2}-m_{4}^{2}}+\cdots ,
\end{eqnarray}%
where $m_{4}$ and $f_{4}$ are the mass and decay constant of the meson $\chi
_{c0}$. The correlator $\Pi ^{\mathrm{OPE}}(p,p^{\prime })$ has the form%
\begin{eqnarray}
&&\Pi ^{\mathrm{OPE}}(p,p^{\prime })=2\int d^{4}xd^{4}ye^{ip^{\prime
}y}e^{-ipx}  \notag \\
&&\times \left\{ \mathrm{Tr}\left[ S_{c}^{ia}(y-x)\widetilde{S}%
_{c}^{jb}(-x){}\widetilde{S}_{c}^{bj}(x)S_{c}^{ai}(x-y)\right] \right.
\notag \\
&&\left. +\mathrm{Tr}\left[ S_{c}^{ib}(y-x)\widetilde{S}_{c}^{ja}(-x){}%
\widetilde{S}_{c}^{bj}(x)S_{c}^{ai}(x-y)\right] \right\} .
\end{eqnarray}%
The following operations are performed in the context of the standard
approach.

In numerical computations, the parameters $M_{2}^{2}$ and $s_{0}^{\prime }$
in the $\chi _{c0}$ channel are chosen in the form
\begin{equation}
M_{2}^{2}\in \lbrack 4,5]~\mathrm{GeV}^{2},\ s_{0}^{\prime }\in \lbrack
14,14.9]~\mathrm{GeV}^{2},
\end{equation}%
where $s_{0}^{\prime }$ is limited by the mass of the charmonium $\chi
_{c0}(3860)$. The coupling $g_{4}$ which corresponds to the vertex $T_{%
\mathrm{4c}}\chi _{c0}\chi _{c0}$ is extracted at $Q^{2}=-m_{4}^{2}$ of the
fit function $\mathcal{G}_{4}(Q^{2})$ with parameters $\mathcal{G}%
_{4}^{0}=0.56$, $c_{4}^{1}=2.81$, and $c_{4}^{2}=-2.98$.

The strong coupling $g_{4}$ equals to
\begin{equation}
g_{4}\equiv \mathcal{G}_{4}(-m_{4}^{2})=(2.4\pm 0.45)\times 10^{-1}\ \mathrm{%
GeV}^{-1}.
\end{equation}%
The partial width of the decay $T_{\mathrm{4c}}\rightarrow \chi _{c0}\chi
_{c0}$ is computed by employing the expression%
\begin{equation}
\Gamma \left[ T_{\mathrm{4c}}\rightarrow \chi _{c0}\chi _{c0}\right]
=g_{4}^{2}\frac{m_{4}^{2}\lambda _{4}}{8\pi }\left( 1+\frac{\lambda _{4}^{2}%
}{m_{4}^{2}}\right) ,
\end{equation}%
where $\lambda _{4}=\lambda (m,m_{4},m_{4})$. Numerical analyses lead to the
result%
\begin{equation}
\Gamma \left[ T_{\mathrm{4c}}\rightarrow \chi _{c0}\chi _{c0}\right] =(16\pm
5)~\mathrm{MeV}.
\end{equation}

The partial widths of six decays of the tetraquark $T_{\mathrm{4c}}$ are
collected in Table\ \ref{tab:Channels}. Using these predictions, it is easy
to find that
\begin{equation}
\Gamma _{\mathrm{4c}}=(128\pm 22)~\mathrm{MeV}.  \label{eq:FW}
\end{equation}

\begin{table}[tbp]
\begin{tabular}{|c|c|c|c|}
\hline\hline
$i$ & Channels & $g_{i}^{(\ast)}\times10~(\mathrm{GeV}^{-1})$ & $%
\Gamma_{i}^{(\ast)}~(\mathrm{MeV})$ \\ \hline
$1$ & $T_{\mathrm{4c}}\to J/\psi J/\psi$ & $5.1 \pm 1.1$ & $56 \pm 18$ \\
$1^{\ast}$ & $T_{\mathrm{4c}}\to J/\psi\psi^{\prime}$ & $4.2 \pm 1.0$ & $15
\pm 5$ \\
$2$ & $T_{\mathrm{4c}}\to \eta_{c}\eta_{c}$ & $1.7 \pm 0.4$ & $24 \pm 8$ \\
$2^{\ast}$ & $T_{\mathrm{4c}}\to \eta_{c}\eta_{c}(2S)$ & $0.9 \pm 0.28 $ & $%
5 \pm 2$ \\
$3$ & $T_{\mathrm{4c}} \to \eta_{c}\chi_{c1}(1P)$ & $25 \pm 6^{\star}$ & $12
\pm 4 $ \\
$4$ & $T_{\mathrm{4c}} \to \chi_{c0}\chi_{c0}$ & $2.4 \pm 0.45 $ & $16 \pm 5
$ \\ \hline
$5$ & $T_{\mathrm{4b}}\to \eta_{b}\eta_{b}$ & $1.9 \pm 0.4 $ & $94 \pm 28$
\\ \hline\hline
\end{tabular}%
\caption{Decay channels of the tetraquarks $T_{\mathrm{4c}}$ and $T_{\mathrm{%
4b}}$, strong couplings $g_{i}^{(\ast)}$, and partial widths $%
\Gamma_{i}^{(\ast)}$. The coupling $g_3$ is dimensionless.}
\label{tab:Channels}
\end{table}

\section{Width of the tetraquark $T_{\mathrm{4b}}$}

\label{sec:Decays3}

In this section, we evaluate the width of the fully heavy tetraquark $T_{%
\mathrm{4b}}$. Our analysis shows that $T_{\mathrm{4b}}$ can dissociate to
mesons $\eta _{b}\eta _{b}$. Investigation of the decay $T_{\mathrm{4b}%
}\rightarrow \eta _{b}\eta _{b}$ can be performed in accordance with a
scheme utilized in the section \ref{sec:Decays2} while considering the
process $T_{\mathrm{4c}}\rightarrow \eta _{c}\eta _{c}$.

The correlation function necessary to extract a sum rule for the form factor
$g_{4}(q^{2})$ in this case is given by the expression
\begin{eqnarray}
&&\Pi _{b}(p,p^{\prime })=i^{2}\int d^{4}xd^{4}ye^{ip^{\prime
}y}e^{-ipx}\langle 0|\mathcal{T}\{J^{\eta _{b}}(y)  \notag \\
&&\times J^{\eta _{b}}(0)J^{\dagger }(x)\}|0\rangle ,  \label{eq:CF10}
\end{eqnarray}%
where $J^{\eta _{b}}(x)$ in the interpolating current for the meson $\eta
_{b}$
\begin{equation}
J^{\eta _{b}}(x)=\overline{b}_{j}(x)i\gamma _{5}b_{j}(x).  \label{eq:C3}
\end{equation}

To determine $g_{5}$, we use the standard "ground-state+continuum" scheme.
Then, it is not difficult to find the physical side of the sum rule, which
is given by the following expression
\begin{eqnarray}
&&\Pi _{b}^{\mathrm{Phys}}(p,p^{\prime })=\frac{\langle 0|J^{\eta _{b}}|\eta
_{b}(p^{\prime })\rangle }{p^{\prime 2}-m_{5}^{2}}\frac{\langle 0|J^{\eta
_{b}}|\eta _{b}(q)\rangle }{q^{2}-m_{5}^{2}}  \notag \\
&&\times \langle \eta _{b}(p^{\prime })\eta _{b}(q)|T_{\mathrm{4b}%
}(p)\rangle \frac{\langle T_{\mathrm{4b}}(p)|J^{\dagger }|0\rangle }{%
p^{2}-m^{2}}+\cdots ,  \label{eq:CF10a}
\end{eqnarray}%
where $m_{5}$ is the mass of the $\eta _{b}$ meson. The matrix elements
which are required to simplify $\Pi _{b}^{\mathrm{Phys}}(p,p^{\prime })$
have the forms
\begin{eqnarray}
&&\langle 0|J^{\eta _{b}}|\eta _{b}\rangle =\frac{f_{5}m_{5}^{2}}{2m_{b}},
\notag \\
&&\langle \eta _{b}(p^{\prime })\eta _{b}(q)|T_{\mathrm{4b}}(p)\rangle
=g_{5}(q^{2})p\cdot p,
\end{eqnarray}%
where $f_{5}$ is the decay constant of $\eta _{b}$.

\begin{figure}[h]
\includegraphics[width=8.5cm]{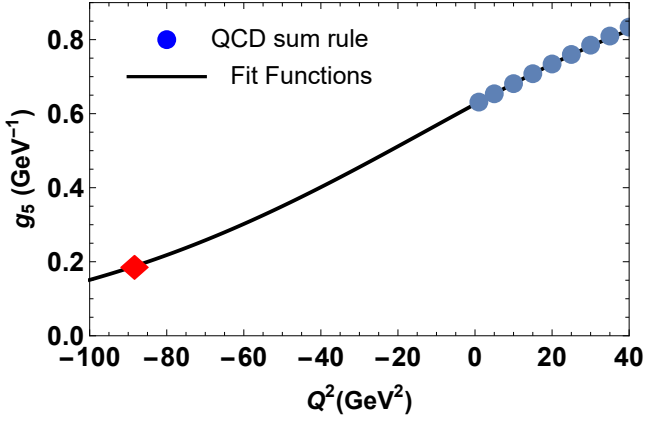}
\caption{The sum rule predictions and fit function for the strong coupling $%
g_{5}(Q^{2})$.}
\label{fig:Fit1}
\end{figure}

To find the QCD side of the sum rule $\Pi ^{\mathrm{OPE}}(p,p^{\prime })$
one needs to replace all $c$-quark propagators in Eq.\ (\ref{eq:QCDside2})
by relevant $S_{b}(x)$ propagators. The sum rule for the coupling predicts%
\begin{equation}
g_{5}\equiv \mathcal{G}_{5}(-m_{5}^{2})=(1.9\pm 0.4)\times 10^{-1}\ \mathrm{%
GeV}^{-1}.
\end{equation}%
The coupling $g_{5}$ is calculated by means of the extrapolating function $%
\mathcal{G}_{5}(Q^{2})$ with $\mathcal{G}_{5}^{0}=0.63$, $c_{5}^{1}=3.21$,
and $c_{5}^{2}=-6.58$. In sum rule computations the parameters $M_{1}^{2}$
and $s_{0}$ in $T_{\mathrm{4b}}$ channel are chosen in accordance with Eq.\ (%
\ref{eq:Wind2}), whereas for $M_{2}^{2}$ and $s_{0}^{\prime }$ in the $\eta
_{b}$ channel, we employ
\begin{equation}
M_{2}^{2}\in \lbrack 10,12]~\mathrm{GeV}^{2},\ s_{0}^{\prime }\in \lbrack
93,97]~\mathrm{GeV}^{2}.
\end{equation}%
Results obtained for $g_{5}(Q^{2})$ and the function $\mathcal{G}_{5}(Q^{2})$
are plotted in Fig.\ \ref{fig:Fit1}.

The width of the decay $T_{\mathrm{4b}}\rightarrow \eta _{b}\eta _{b}$ can
be found by means of the expression Eq.\ (\ref{eq:PDw2}) with evident
substitutions. Our prediction reads
\begin{equation}
\Gamma \left[ T_{\mathrm{4b}}\rightarrow \eta _{b}\eta _{b}\right] =(94\pm
28)~\mathrm{MeV}.
\end{equation}

\section{Discussion and concluding notes}

\label{sec:Disc}

We have explored the fully charmed and beauty tetraquarks $T_{\mathrm{4c}}$
and $T_{\mathrm{4b}}$ in the context of QCD sum rule method. We have modeled
them as diquark-antidiquark systems composed of $^{3}\mathrm{P}_{0}$
pseudoscalar states $Q^{T}CQ$ and $\overline{Q}C\overline{Q}^{T}$ with
quantum numbers $J^{\mathrm{P}}=0^{-}$. Our present investigations include
detailed calculations of both the mass and width of these tetraquarks.

Predictions obtained for the mass and full width of $T_{\mathrm{4c}}$
\begin{eqnarray}
&&m=(6928\pm 50)~\mathrm{MeV}  \notag \\
&&\Gamma _{\mathrm{4c}}=(128\pm 22)~\mathrm{MeV}.
\end{eqnarray}%
allow us to confront them with the available data of the LHCb-ATLAS-CMS
Collaborations. The data of interest, reported by these experiments are
\begin{eqnarray}
&&m^{\mathrm{LHCb}} =(6905\pm 11\pm 7)~\mathrm{MeV},  \notag \\
&&\Gamma ^{\mathrm{LHCb}} =(80\pm 19\pm 33)~\mathrm{MeV},  \label{eq:MW1}
\end{eqnarray}

\begin{eqnarray}
&&m^{\mathrm{ATL}}=6870\pm 30_{-10}^{+60}~\mathrm{MeV},  \notag \\
&&\Gamma ^{\mathrm{ATL}}=120\pm 40_{-10}^{+30}~\mathrm{MeV},
\label{eq:MWATL3}
\end{eqnarray}%
and%
\begin{eqnarray}
&&m^{\mathrm{CMS}}=(6927\pm 9\pm 5)~\mathrm{MeV},  \notag \\
&&\Gamma ^{\mathrm{CMS}}=(122\pm 22\pm 19)~\mathrm{MeV},  \label{eq:MWCMS2}
\end{eqnarray}%
respectively. It is evident that $m$ and $\Gamma _{\mathrm{4c}}$ are in
excellent agreements with $m^{\mathrm{CMS}}$ and $\Gamma ^{\mathrm{CMS}}$ of
the CMS Collaboration. Within errors of calculations, they are also
compatible with data of other experiments. Therefore, by taking into account
these circumstances, we have inclined to idea that this tetraquark is nice
candidate to the resonance $X(6900)$. Our conclusions about the nature of $%
X(6900)$ are in accord with those of Ref.\ \cite{Albuquerque:2020hio}.

In Ref.\ \cite{Agaev:2023wua}, we performed an analysis for the fully
charmed scalar diquark-antidiquark state $X_{\mathrm{4c}}$ built of $^{3}%
\mathrm{S}_{1}$ states $Q^{T}C\gamma _{\mu }Q$ and $\overline{Q}\gamma _{\mu
}C\overline{Q}^{T}$ with spin-parity $J^{\mathrm{P}}=1^{+}$. It is known,
that different diquark-andiquark currents can be expressed in terms of
molecular-type currents by means of the Fierz transformations \cite%
{Chen:2022sbf,Xin:2021wcr}. Contrary, a molecule current can be presented as
a weighted sum of different diquark-antidiquark ones \cite{Wang:2020rcx}. By
performing similar operations, it is possible to find some common pieces in
the currents used in Ref.\ \cite{Agaev:2023wua} and in the present work.
However, four-quark systems composed of $^{3}\mathrm{P}_{0}$ or $^{3}\mathrm{%
S}_{1}$ diquarks have different inner organizations and parameters. For
instance, the mass $(6570\pm 55)~\mathrm{MeV}$ of the tetraquark $X_{\mathrm{%
4c}}$ is considerably smaller than $m$, which allowed us to identify it in
Ref.\ \cite{Agaev:2023wua} with the resonance $X(6600)$.

The fully beauty exotic mesons during past years were also under intensive
investigations. The tetraquark $T_{\mathrm{4b}}$ that has been studied in
this article, is a beauty counterpart of the fully charmed state $T_{\mathrm{%
4c}}$. It turned out that $T_{\mathrm{4b}}$ can decay to $\eta _{b}\eta _{b}$
pairs and be detected in a mass distribution of these mesons. Its parameters
may be interesting for future experimental studies of fully beauty
resonances.

\begin{widetext}

\appendix*

\section{ Heavy quark propagator $S_{Q}^{ab}(x)$ and spectral density $%
\protect\rho ^{\mathrm{pert.}}(s,\protect\alpha ,\protect\beta ,\protect%
\gamma )$}

\renewcommand{\theequation}{\Alph{section}.\arabic{equation}} \label{sec:App}

In this paper, we use for the heavy quark propagator $S_{Q}^{ab}(x)$ ($Q=c,\
b$) the following expression
\begin{eqnarray}
&&S_{Q}^{ab}(x)=i\int \frac{d^{4}k}{(2\pi )^{4}}e^{-ikx}\Bigg \{\frac{\delta
_{ab}\left( {\slashed k}+m_{Q}\right) }{k^{2}-m_{Q}^{2}}-\frac{%
g_{s}G_{ab}^{\alpha \beta }}{4}\frac{\sigma _{\alpha \beta }\left( {\slashed %
k}+m_{Q}\right) +\left( {\slashed k}+m_{Q}\right) \sigma _{\alpha \beta }}{%
(k^{2}-m_{Q}^{2})^{2}}  \notag \\
&&+\frac{g_{s}^{2}G^{2}}{12}\delta _{ab}m_{Q}\frac{k^{2}+m_{Q}{\slashed k}}{%
(k^{2}-m_{Q}^{2})^{4}}+\cdots \Bigg \},
\end{eqnarray}%
where the notations
\begin{equation}
G_{ab}^{\alpha \beta }\equiv G_{A}^{\alpha \beta }\lambda _{ab}^{A}/2,\ \
G^{2}=G_{\alpha \beta }^{A}G_{A}^{\alpha \beta },\
\end{equation}%
have been employed. Here, $G_{A}^{\alpha \beta }$ is the gluon
field-strength tensor, and $\lambda ^{A}$ are the Gell-Mann matrices. The
indices $A,B,C$ run in the range $1,2,\ldots 8$.

The invariant amplitude $\Pi (M^{2},s_{0})$ obtained after the Borel
transformation and continuum subtraction has the form
\begin{equation*}
\Pi (M^{2},s_{0})=\int_{16m_{Q}^{2}}^{s_{0}}ds\rho ^{\mathrm{OPE}%
}(s)e^{-s/M^{2}},
\end{equation*}%
where the spectral density $\rho ^{\mathrm{OPE}}(s)$ is determined by the
formula
\begin{equation*}
\rho ^{\mathrm{OPE}}(s)=\rho ^{\mathrm{pert.}}(s)+\langle \alpha
_{s}G^{2}/\pi \rangle \rho ^{\mathrm{Dim4}}(s).
\end{equation*}%
The components $\rho ^{\mathrm{pert.}}(s)$ and $\rho ^{\mathrm{Dim4}}(s)$ of
the spectral density are
\begin{equation}
\rho ^{\mathrm{pert.(Dim4)}}(s)=\int_{0}^{1}d\alpha \int_{0}^{1-a}d\beta
\int_{0}^{1-a-\beta }d\gamma \rho ^{\mathrm{pert.(Dim4)}}(s,\alpha ,\beta
,\gamma ),  \label{eq:A2}
\end{equation}%
with $\alpha $, $\beta $, and $\gamma $ being the Feynman parameters.

The function $\rho ^{\mathrm{pert.}}(s,\alpha ,\beta ,\gamma )$ is given by
the formula
\begin{eqnarray}
&&\rho ^{\mathrm{pert.}}(s,\alpha ,\beta ,\gamma )=\frac{\Theta
(L_{3})N_{1}^{2}}{2048\pi ^{6}N_{2}^{8}N_{3}^{5}L_{1}^{2}}%
\{33N_{1}^{2}(L\gamma N_{3}^{2}L\gamma +N_{3}(-2\gamma ^{4}+\alpha
^{2}L_{1}(-2+3(\gamma +\beta )+4\alpha )-L_{1}^{2}(\beta \gamma +\alpha
(\beta +\gamma ))  \notag \\
&&+\alpha (\gamma ^{4}(\alpha +\beta )+\alpha L_{2}^{2}(\beta (\beta
-1)+\alpha (\beta -1)+\alpha ^{2})+\gamma ^{3}(3\beta (\beta -1)+\alpha
(5\beta -3)+3\alpha ^{2})+\gamma ^{2}L_{2}(3\beta (\beta -1)  \notag \\
&&+\alpha (5\beta -3)+4\alpha ^{2})+\gamma L_{2}(\beta (\beta -1)^{2}+\alpha
+\alpha \beta (4\beta -5)+\alpha ^{2}(5\beta -4)+3\alpha
^{2})))+6(11L^{5}s^{2}\gamma \alpha ^{5}(L\gamma \alpha -N_{2})^{3}  \notag
\\
&&+N_{3}L^{2}s\alpha ^{2}(L\gamma \alpha -N_{2})^{2}(22N_{2}L^{2}s\gamma
\alpha ^{2}-22L^{3}s\gamma ^{2}\alpha
^{3}+N_{2}^{2}m_{Q}^{2}(L_{1}^{2}+11L_{1}\alpha -11\alpha
^{2}))+N_{2}^{2}N_{3}^{3}m_{Q}^{2}(11N_{2}^{2}(m_{Q}^{2}L_{1}^{2}-Ls\alpha )
\notag \\
&&-N_{2}Ls\gamma \alpha (L_{1}^{2}+22L_{1}\alpha -22\alpha
^{2})+L^{2}s\gamma ^{2}\alpha ^{2}(L_{1}^{2}+11L_{1}\alpha -11\alpha
^{2}))+N_{3}^{2}Ls\alpha (L\gamma \alpha -N_{2})(-22N_{2}L^{3}s\gamma
^{2}\alpha ^{3}+11L^{4}s\gamma ^{3}\alpha ^{4}  \notag \\
&&+N_{2}^{3}m_{Q}^{2}(L_{1}^{2}+22L_{1}\alpha -22\alpha
^{2})+N_{2}^{2}L\gamma \alpha (11Ls\alpha +m_{Q}^{2}\left(
-13L_{1}^{2}-22L_{1}\alpha +22\alpha ^{2}\right)
)))-2N_{1}(2N_{3}^{3}N_{2}^{2}m_{Q}^{2}(L_{1}^{2}+11L_{1}\alpha  \notag \\
&&-11\alpha ^{2})+N_{3}(-132L^{5}s\gamma ^{2}\alpha ^{5}+33N_{2}L^{3}s\gamma
\alpha ^{3}(L_{1}\gamma +6\alpha (L_{2}+\gamma ))+N_{2}^{2}(-33L^{2}s\alpha
^{2}(L_{1}\gamma +2\alpha (L_{2}+\gamma ))+2m_{Q}^{2}(L_{1}^{2}  \notag \\
&&+11L_{1}\alpha -11\alpha ^{2})(-\alpha ^{2}+L_{1}(\alpha +\beta ))(-\alpha
^{2}+L_{1}(\alpha +\gamma ))))+33L^{3}s\alpha ^{3}(2L^{3}\alpha ^{3}\gamma
^{2}+N_{2}(L_{1}\gamma +\alpha (L_{2}+\gamma ))  \notag \\
&&+N_{2}\alpha \gamma (L_{1}^{2}\gamma -\alpha (3L_{2}^{2}+\gamma
(-5+2\gamma +5\beta +6\alpha ))))+N_{3}^{2}(-99N_{2}L^{3}s\gamma \alpha
^{3}+66L^{4}s\gamma ^{2}\alpha ^{4}+N_{2}^{2}(33L^{2}s\alpha
^{2}+2m_{Q}^{2}(22\alpha ^{4}  \notag \\
&&-L_{1}\alpha ^{2}\left( -9+20\gamma +20\beta +44\alpha \right)
+L_{1}^{2}(11\gamma \beta +13\alpha -2(\beta +\gamma )\alpha )))))\}.
\end{eqnarray}%
In expression above, $\Theta (z)$ is the Unit Step function. We have
introduced also the following notations%
\begin{eqnarray}
&&N_{1}=s\alpha \beta \gamma \left[ \gamma ^{3}+\alpha (\beta +\alpha
-1)^{2}+\gamma (\beta +\alpha -1)(-1+2\gamma +\alpha +2\beta )\right]
-m_{Q}^{2}\left[ \gamma ^{4}(\beta +\alpha )+\alpha (\beta +\alpha
-1)^{2}\right.  \notag \\
&&\left. \times (2\gamma \beta +(\gamma +\beta )\alpha )+\gamma ^{2}(\beta
+\alpha -1)(\beta ^{2}-\alpha +2\alpha (\alpha +\gamma )+\beta (-1+2\gamma
+\alpha +4\beta ))\right] ,  \notag \\
&&N_{2}=\beta \alpha (\alpha +\beta -1)+\gamma ^{2}(\alpha +\beta )+\gamma
\left[ \beta ^{2}+\alpha (\alpha -1)+\beta (2\alpha -1)\right] ,\
N_{3}=\gamma ^{2}+(\gamma +\alpha )(\beta +\alpha -1),  \notag \\
&&L =\alpha +\beta +\gamma -1,\ L_{1}=1-\beta -\gamma ,\ L_{2}=\alpha +\beta
-1,\ L_{3}=N_{3}\left[ s\alpha \beta \gamma L-m_{Q}^{2}N_{2}\right]
/N_{2}^{2}.
\end{eqnarray}

\end{widetext}


\begin{thebibliography}{999}

\bibitem{Jaffe:1976ig} R.~L.~Jaffe, \textit{Multi-Quark Hadrons. 1. The
Phenomenology of }$q^{2}\overline{q}^{2}$\textit{\ Mesons}, Phys.\ Rev.\ D
\textbf{15}, 267 (1977). 


\bibitem{Jaffe:1976yi} R.~L.~Jaffe, \textit{Perhaps a Stable Dihyperon},
Phys.\ Rev.\ Lett.\ \textbf{38}, 195 (1977); \textbf{38}, 617(E) (1977).


\bibitem{Ader:1981db} J.~P.~Ader, J.~M.~Richard, and P.~Taxil, \textit{Do
Narrow Heavy Multi - Quark States Exist?}, Phys.\ Rev.\ D \textbf{25}, 2370
(1982).


\bibitem{Zouzou:1986qh} S.~Zouzou, B.~Silvestre-Brac, C.~Gignoux, and
J.~M.~Richard, \textit{Four Quark Bound States}, Z.\ Phys.\ C \textbf{30},
457 (1986).


\bibitem{Lipkin:1986dw} H.~J.~Lipkin, \textit{A Model Independent Approach
To Multi - Quark Bound States}, Phys.\ Lett.\ B \textbf{172}, 242 (1986).


\bibitem{Carlson:1987hh} J.~Carlson, L.~Heller, and J.~A.~Tjon, \textit{%
Stability of Dimesons}, Phys.\ Rev.\ D \textbf{37}, 744 (1988).


\bibitem{Navarra:2007yw} F.~S.~Navarra, M.~Nielsen, and S.~H.~Lee, \textit{%
QCD sum rules study of }$QQ\bar{u}\bar{d}$ \textit{mesons}, Phys.\ Lett.\ B
\textbf{649}, 166 (2007).


\bibitem{Eichten:2017ffp} E.~J.~Eichten and C.~Quigg, \textit{Heavy-quark
symmetry implies stable heavy tetraquark mesons }$Q_{i}Q_{j}\bar{q}_{k}\bar{q%
}_{l}$, Phys.\ Rev.\ Lett.\ \textbf{119}, 202002 (2017).


\bibitem{Karliner:2017qjm} M.~Karliner and J.~L.~Rosner, \textit{Discovery
of doubly-charmed }$\Xi _{cc}$\textit{\ baryon implies a stable (}$bb\bar{u}%
\bar{d}$\textit{) tetraquark}, Phys.\ Rev.\ Lett.\ \textbf{119}, 202001
(2017).


\bibitem{Xing:2018bqt} Y.~Xing and R.~Zhu, \textit{Weak decays of the stable
doubly heavy tetraquark states}, Phys.\ Rev.\ D \textbf{98}, 053005 (2018).


\bibitem{Agaev:2018khe} S.~S.~Agaev, K.~Azizi, B.~Barsbay, and H.~Sundu,
\textit{Weak decays of the axial-vector tetraquark }$T_{bb;\bar{u}\bar{d}%
}^{-}$, Phys.\ Rev.\ D \textbf{99}, 033002 (2019).


\bibitem{Li:2018bkh} G.~Li, X.~F.~Wang and Y.~Xing, \textit{SU(3) analysis
of weak decays of doubly-heavy tetraquarks }$b\overline{c}q\overline{q}$,
Eur.\ Phys.\ J.\ C \textbf{79}, 210 (2019). 


\bibitem{Sundu:2019feu} H.~Sundu, S.~S.~Agaev, and K.~Azizi, \textit{%
Semileptonic decays of the scalar tetraquark }$Z_{bc;\overline{u}\overline{d}%
}^{0}$, Eur.\ Phys.\ J.\ C \textbf{79}, 753 (2019).


\bibitem{Agaev:2019kkz} S.~S.~Agaev, K.~Azizi, and H.~Sundu, \textit{%
Double-heavy axial-vector tetraquark }$T_{bc;\bar{u}\bar{d}}^{0}$, Nucl.\
Phys.\ B \textbf{951}, 114890 (2020). 


\bibitem{Agaev:2019lwh} S.~S.~Agaev, K.~Azizi, B.~Barsbay, and H.~Sundu,
\textit{Heavy exotic scalar meson }$T_{bb;\bar{u}\bar{s}}^{-}$, Phys.\ Rev.\
D \textbf{101}, 094026 (2020). 


\bibitem{Agaev:2020dba} S.~S.~Agaev, K.~Azizi, B.~Barsbay, and H.~Sundu,
\textit{Stable scalar tetraquark }$T_{bb;\bar{u}\bar{d}}^{-}$, Eur.\ Phys.\
J.\ A \textbf{56}, 177 (2020). 


\bibitem{Agaev:2020mqq} S.~S.~Agaev, K.~Azizi, B.~Barsbay, and H.~Sundu,
\textit{Semileptonic and nonleptonic decays of the axial-vector tetraquark }$%
T_{bb;\bar{u}\bar{d}}^{-}$, Eur.\ Phys.\ J.\ A \textbf{57}, 106 (2021).


\bibitem{Agaev:2020zag} S.~S.~Agaev, K.~Azizi, B.~Barsbay, and H.~Sundu,
\textit{A family of double-beauty tetraquarks: Axial-vector state }$T_{bb;%
\bar{u}\bar{s}}^{-}$, Chin.\ Phys.\ C \textbf{45}, 013105 (2021).


\bibitem{Yu:2017pmn} F.~S.~Yu, \textit{Weak-decay searches for }$Qs\overline{%
u}\overline{d}$, Eur.\ Phys.\ J.\ C \textbf{82}, 641 (2022).


\bibitem{LHCb:2020bwg} R.~Aaij \textit{et al.} (LHCb Collaboration), \textit{%
Observation of structure in the }$J/\psi $\textit{\ -pair mass spectrum},
Sci.\ Bull. \textbf{65}, 1983 (2020).


\bibitem{Bouhova-Thacker:2022vnt} E.~Bouhova-Thacker (ATLAS Collaboration),
\textit{ATLAs results on exotic hadronic resonances}, PoS \textbf{ICHEP2022}%
, 806 (2022).


\bibitem{CMS:2023owd} A.~Hayrapetyan, \textit{et al.} (CMS Collaboration),
\textit{Recent CMS results on exotic resonances}, arXiv:2306.07164 [hep-ex].


\bibitem{Zhang:2020xtb} J.~R.~Zhang, $0^{+}$\textit{\ fully-charmed
tetraquark states}, Phys.\ Rev.\ D \textbf{103}, 014018 (2021).


\bibitem{Albuquerque:2020hio} R.~M.~Albuquerque, S.~Narison,
A.~Rabemananjara, D.~Rabetiarivony, and G.~Randriamanatrika, \textit{%
Doubly-hidden scalar heavy molecules and tetraquarks states from QCD at NLO}%
, Phys.\ Rev.\ D \textbf{102}, 094001 (2020).


\bibitem{Wang:2022xja} Z.~G.~Wang, \textit{Analysis of the }$X(6600)$\textit{%
, }$X(6600)$, $X(7300)$\textit{\ and related tetraquark states with the QCD
sum rules}, Nucl.\ Phys.\ B \textbf{985}, 115983 (2022).


\bibitem{Dong:2022sef} W.~C.~Dong and Z.~G.~Wang, \textit{Going in quest of
potential tetraquark interpretaions for the newly observed $T_{\psi\psi}$
states in light of the diquark-antidiquark scenarios}, Phys.\ Rev.\ D
\textbf{107}, 074010 (2023). 


\bibitem{Faustov:2022mvs} R.~N.~Faustov, V.~O.~Galkin, and E.~M.~Savchenko,
\textit{Fully-heavy tetraquark spectroscopy in the relativistic quark model}%
, Symmetry \textbf{14}, 2504 (2022). 


\bibitem{Lu:2023ccs} Y.~Lu, C.~Cheng, K.~G.~Kang, G.~y.~Qin, and
H.~Q.~Zheng, $X(6900)$\textit{\ peak could be a molecular state}, Phys.\
Rev.\ D \textbf{107}, 094006 (2023). 


\bibitem{Dong:2020nwy} X.~K.~Dong, V.~Baru, F.~K.~Guo, C.~Hanhart, and
A.~Nefediev, \textit{Coupled-Channel Interpretation of the LHCb Double-~}$%
J/\psi $\textit{~Spectrum and Hints of a New State Near the~ }$J/\psi J/\psi
$\textit{~~Threshold}, Phys.\ Rev.\ Lett. \textbf{126}, 132001 (2021);
\textbf{127}, 119901(E) (2021).


\bibitem{Liang:2021fzr} Z.~R.~Liang, X.~Y.~Wu, and D.~L.~Yao, \textit{%
Hunting for states in the recent LHCb di-}$J/\psi $ \textit{invariant mass
spectrum}, Phys.\ Rev.\ D \textbf{104}, 034034 (2021).


\bibitem{Gong:2020bmg} C.~Gong, M.~C.~Du, Q.~Zhao, X.~H.~Zhong, and B.~Zhou
\textit{Nature of }$X(6900)$\textit{\ and its production mechanism at LHCb},
Phys.\ Lett.\ B \textbf{824}, 136794 (2022). 


\bibitem{Niu:2022vqp} P.~Niu, Z.~Zhang, Q.~Wang, and M.~L.~Du, \textit{The
third peak structure in the double }$J/\psi $\textit{\ spectrum},
arXiv:2212.06535.


\bibitem{Yu:2022lak} G.~L.~Yu, Z.~Y.~Li, Z.~G.~Wang, J.~Lu, and M.~Yan,
\textit{The }$S$\textit{\ and }$P$\textit{\ wave fully charmed tetraquark
states and their radial excitations}, Eur.\ Phys.\ J. C \textbf{83}, 416
(2023). 


\bibitem{Kuang:2023vac} S.~Q.~Kuang, Q.~Zhou, D.~Guo, Q.~H.~Yang, and
L.~Y.~Dai, \textit{Study of }$X(6900)$\textit{\ with unitarized coupled
channel scattering amplitudes}, Eur.\ Phys.\ J. C \textbf{83}, 383 (2023).


\bibitem{Agaev:2023wua} S.~S.~Agaev, K.~Azizi, B.~Barsbay, and H.~Sundu,
Phys.\ Lett.\ B \textbf{844}, 138089 (2023). 


\bibitem{Becchi:2020mjz} C.~Becchi, A.~Giachino, L.~Maiani, and
E.~Santopinto, \textit{Search for }$bb\bar{b}\bar{b}$\textit{\ tetraquark
decays in 4 muons, }$B^{+}B^{-}$, $B^{0}\bar{B}^{0}$\textit{\ and }$B_{s}^{0}%
\bar{B}_{s}^{0}$\textit{\ channels at LHC}, Phys.\ Lett.\ B \textbf{806},
135495 (2020).


\bibitem{Becchi:2020uvq} C.~Becchi, A.~Giachino, L.~Maiani, and
E.~Santopinto, \textit{A study of }$cc\bar{c}\bar{c}$\textit{\ tetraquark
decays in 4 muons and in }$D^{(\ast )}\bar{D}^{(\ast )}$\textit{\ at LHCb},
Phys.\ Lett.\ B \textbf{811}, 135952 (2020). 


\bibitem{Shifman:1978bx} M.~A.~Shifman, A.~I.~Vainshtein and V.~I.~Zakharov,
\textit{QCD and Resonance Physics. Theoretical Foundations}, Nucl.\ Phys.\ B
\textbf{147}, 385 (1979).


\bibitem{Shifman:1978by} M.~A.~Shifman, A.~I.~Vainshtein and V.~I.~Zakharov,
\textit{QCD and Resonance Physics: Applications}, Nucl.\ Phys.\ B \textbf{147%
}, 448 (1979).


\bibitem{Agaev:2020zad} S.~S.~Agaev, K.~Azizi, and H.~Sundu, \textit{%
Four-quark exotic mesons}, Turk.\ J.\ Phys.\ \textbf{44}, 95 (2020).


\bibitem{PDG:2022} R.~L.~Workman \textit{et al.} (Particle Data Group),
Prog.\ Theor.\ Exp.\ Phys.\ \textbf{2022}, 083C01 (2022).


\bibitem{Barnes:2005pb} T.~Barnes, S.~Godfrey, and E.~S.~Swanson, \textit{%
Higher charmonia}, Phys.\ Rev.\ D \textbf{72}, 054026 (2005).


\bibitem{Kiselev:2001xa} V.~V.~Kiselev, A.~K.~Likhoded, O.~N.~Pakhomova, and
V.~A.~Saleev, \textit{Leptonic constants of heavy quarkonia in potential
approach of NRQCD}, Phys.\ Rev.\ D \textbf{65}, 034013 (2002).


\bibitem{Hatton:2020qhk} D.~Hatton \textit{et al.} (HPQCD Collaboration),
\textit{Charmonium properties from lattice QCD+QED: hyperfine splitting, }$%
J/\psi $\textit{\ leptonic width, charm quark mass and }$a_{\mu }^{c}$,
Phys.\ Rev.\ D \textbf{102}, 054511 (2020). 


\bibitem{Hatton:2021dvg} D.~Hatton, C.~T.~H.~Davies, J.~Koponen,
G.~P.~Lepage, and A.~T.~Lytle, \textit{Bottomonium precision tests from full
lattice QCD: hyperfine splitting, }$\Upsilon $\textit{\ leptonic width and} $%
b$\textit{\ quark contribution to }$e^{+}e^{-}\rightarrow $\textit{\ hadrons}%
, Phys.\ Rev.\ D \textbf{103}, 054512 (2021). 


\bibitem{VeliVeliev:2012cc} E.~Veli Veliev, K.~Azizi, H.~Sundu, and G.~Kaya,
\textit{Spectrum of the heavy axial-vector }$\chi _{b1}(1P)$\textit{\ and }$%
\chi _{c1}(1P)$, PoS (Confinement X) 339, 2012; arXiv:1205.5703.


\bibitem{Veliev:2010gb} E.~V. Veliev, H.~Sundu, K.~Azizi, and M.~Bayar,
\textit{Scalar quarconia at finite temperature}, Phys.\ Rev.\ D \textbf{82},
056012 (2010). 


\bibitem{Chen:2022sbf} H.~X.~Chen, Y.~X.~Yan, and W.~Chen, \textit{Decay
behaviors of the fully bottom and charm tetraquarks}, Phys.\ Rev.\ D \textbf{%
106}, 094019 (2022). 


\bibitem{Xin:2021wcr} Q.~Xin and Z.~G.~Wang, \textit{Analysis of doubly
charmed tetraquark molecular states with the QCD sum rules,} Eur.\ Phys.\
J.\ A \textbf{58}, 118 (2022). 


\bibitem{Wang:2020rcx} \textit{Q.~N.~Wang, W.~Chen, and H.~X.~Chen, Exotic
molecular states and tetraquark states with }$J^{P}=0^{+}$, Chin.\ Phys.\ C
\textbf{45}, 093102 (2021). 
\end{thebibliography}
\end{document}